\documentclass[a4paper,12pt]{article}
\pdfoutput=1
\usepackage{amsmath,amsfonts,amssymb}
\usepackage{youngtab}
\usepackage{graphicx}
\usepackage{enumerate}
\usepackage{epsfig}
\usepackage{soul}
\usepackage{verbatim,setspace}
\usepackage{xparse}
\usepackage{tikz}
\usepackage{cite}
\usetikzlibrary{calc}

\def\unit{\relax{\rm 1\kern-.26em I}}

\DeclareDocumentCommand{\hcancel}{mO{0pt}O{0pt}O{0pt}O{0pt}}{%
    \tikz[baseline=(tocancel.base)]{
        \node[inner sep=0pt,outer sep=0pt] (tocancel) {#1};
        \draw[black] ($(tocancel.south west)+(#2,#3)$) -- ($(tocancel.north east)+(#4,#5)$);
    }
}

\newcommand{\cD}{{\cal D}}

\newcommand{\cL}{{\cal L}}
\newcommand{\cM}{{\cal M}}

\newcommand{\ov}{\overline}

\newcommand{\Tr}{\mbox{Tr}}
\newcommand{\tr}{\text{tr}}

\newcommand{\be}{\begin{equation}}
\newcommand{\ee}{\end{equation}}
\newcommand{\bea}{\begin{eqnarray}}
\newcommand{\eea}{\end{eqnarray}}

\DeclareMathSymbol{\mg}{\mathrel}{symbols}{"1D}

\newcommand{\MET}{E\llap{/\kern1.5pt}_T}
\newcommand{\nn}{\nonumber}

\def\nn{\nonumber}
\def\bra{\langle}
\def\ket{\rangle}

\def\tr{\mathrm{tr}}
\def\beq{\begin{equation}}
\def\eeq{\end{equation}}

\def\psqr#1#2{{\vcenter{\vbox{\hrule height.#2pt
        \hbox{\vrule width.#2pt height#1pt \kern#1pt
        \vrule width.#2pt}
        \hrule height.#2pt \hrule height.#2pt
        \hbox{\vrule width.#2pt height#1pt \kern#1pt
        \vrule width.#2pt}
        \hrule height.#2pt}}}}
\def\sqr#1#2{{\vcenter{\vbox{\hrule height.#2pt
        \hbox{\vrule width.#2pt height#1pt \kern#1pt
        \vrule width.#2pt}
        \hrule height.#2pt}}}}

\newcommand{\C}[1]{\mathcal{#1}}

\def\ov{\overline}

\newcounter{oldcounter}
\addtocounter{equation}{1}
\begin{document}

\thispagestyle{empty}

\begin{center}
\begin{spacing}{2}
{\LARGE {\bf Dirac Gauginos in Low Scale Supersymmetry Breaking}}
\end{spacing}
\vspace{0.5cm}
{\bf Mark D.~ Goodsell$^{a,b,}$\footnote{mark.goodsell@lpthe.jussieu.fr} and Pantelis Tziveloglou$^{c,d,}$\footnote{pantelis.tziveloglou@vub.ac.be}}

\noindent {\small $^a$Sorbonne Universit\'es, UPMC Univ Paris 06, UMR 7589, LPTHE, F-75005, Paris, France \\
$^b$CNRS, UMR 7589, LPTHE, F-75005, Paris, France \\[0.2cm]
$^c$Theoretische Natuurkunde and IIHE, Vrije Universiteit Brussel,\\ Pleinlaan 2, B-1050 Brussels, Belgium \\[0.2cm]
$^d$International Solvay Institutes,\\ Brussels, Belgium
}
\vspace{0.5cm}

%\today
\end{center}

 \begin{abstract}
\noindent It has been claimed that Dirac gaugino masses are necessary for realistic models of low-scale supersymmetry breaking, and yet very little attention has been paid to the phenomenology of a light gravitino when gauginos have Dirac masses. We begin to address this deficit by investigating the couplings and phenomenology of the gravitino in the effective Lagrangian approach. We pay particular attention to the phenomenology of the scalar octets, where new decay channels open up. This leads us to propose a new simplified effective scenario including only light gluinos, sgluons and gravitinos, allowing the squarks to be heavy -- with the possible exception of the third generation. Finally, we comment on the application of our results to Fake Split Supersymmetry. 
\end{abstract}

\newpage

\tableofcontents

\setcounter{page}{1}
%\twocolumn[]

\newpage

%%%%%%%%%%%%%%%
\section{Introduction}\label{intro}
%%%%%%%%%%%%%%%

The non-observation of supersymmetric particles so far, has cast doubts on our expectations for a solution of the hierarchy problem within the standard realizations of supersymmetry (SUSY). One direction beyond minimality that has received much interest in the recent years is to consider models where the SUSY breaking scale is in the multi-TeV range \cite{Brignole:1996fn,Ponton,Brignole:1997pe,Zwirner,Clark,Brignole:1998me,Brignole:2003cm,Komargodski:2009rz,Antoniadis:2010hs,Dudas:2012fa}. Indeed, although there are increasingly stringent bounds on superpartners from LHC, a \emph{low} scale of SUSY breaking can actually help reduce fine-tuning, as it is associated with a low cutoff for the effective supersymmetric theory so that there are no large effects from renormalisation-group running of SUSY breaking parameters over different energy scales.

Lowering the SUSY breaking scale to the multi-TeV range has several important consequences. First, the gravitino in these models is within the milli-eV range, introducing a new ultralight fermionic degree of freedom in the low energy theory.
According to the high-energy equivalence theorem \cite{cddfg}, at LHC energies gravitino dynamics are dominated by the goldstino longitudinal component. Therefore we can use this spin-1/2 fermionic field to describe gravitino interactions with the rest of the fields. Second, the Higgs sector is modified in a way that alleviates the little hierarchy problem \cite{Brignole:2003cm,Antoniadis:2010hs}. Third, since the cutoff scale is low, higher-dimensional effective operators can introduce interesting phenomenological effects such as contributions to $h\to \gamma\gamma/ \gamma Z$, $gg\to h$, wrong-Higgs Yukawa couplings, monophoton+MET signals and four-femion contact interactions \cite{Brignole:2003cm,Dudas:2012fa}. Fourth, while in principle it is possible to write down an effective theory with Majorana gauginos and SUSY broken at a low scale, in practice it is very difficult to construct an explicit model where the Majorana masses are not unacceptably small. Model building aspects of low-scale SUSY breaking suggest that the gauginos should be of Dirac type \cite{Gherghetta:2011na,Benakli:2007zza}.

This last consequence has many far reaching implications for the phenomenology of low-scale breaking models which have received very little attention so far in the literature, with the exception of \cite{Benakli:2007zza,Benakli:2014daa} (in the context of brane-worlds). On the other hand, conventional signatures of Dirac gauginos are attracting much attention, see for example \cite{fayet,Polchinski:1982an,Hall:1990hq,fnw,Nelson:2002ca,Antoniadis:2005em,Antoniadis:2006uj,kpw,%
Amigo:2008rc,Plehn,Benakli:2008pg,Choi:2008ub,Belanger:2009wf,Benakli:2009mk,Choi:2009ue,Benakli:2010gi,Choi:2010gc,%
Carpenter:2010as,Kribs:2010md,Abel:2011dc,Davies:2011mp,Benakli:2011kz,Kalinowski:2011zz,Frugiuele:2011mh,%
ItoyamaMaru,Rehermann:2011ax,Bertuzzo:2012su,Argurio:2012cd,Goodsell:2012fm,Fok:2012fb,Argurio:2012bi,Frugiuele:2012pe,%
Frugiuele:2012kp,Benakli:2012cy,Kribs:2013oda,Kribs:2013eua,Chakraborty:2013gea,Csaki:2013fla,Beauchesne:2014pra,%
Benakli:2014daa,%
Bertuzzo:2014bwa,Benakli:2014cia,Chakraborty:2014tma}. The supersymmetric multiplets that need to be added in order to make the gauginos Dirac include scalar fields in the adjoint representation of the gauge groups, and the potential of detecting some of these particles, in particular the colour octets, has been extensively investigated in the literature \cite{Atlassgluon,Atlassgluon2,Plehn,Choi:2008ub,Schumann,Fuks} as well as their effect on the scalar potential and the Higgs sector of the theory \cite{Belanger:2009wf,Benakli:2011kz,Frugiuele:2011mh,Benakli:2012cy,Bertuzzo:2014bwa,Benakli:2014cia}. However, in the presence of gravitino interactions we expect that the experimental signatures of the adjoints will be modified. Furthermore, Dirac gauginos have modified interactions with gravitinos compared to the standard interactions of Majorana gauginos. This implies that the limits and studies of such models should be revisited. 

The goal of this paper is to study these effects and highlight the interesting features for future LHC phenomenological studies. In section \ref{SEC:GENERICCOUPLINGS} we use effective field theory to construct the general expression of the couplings of the goldstino to the adjoint multiplets and we discuss the new physics that these couplings introduce with respect to the Majorana case. In section \ref{sgluondecays} we compute the widths of the various decay modes of the colour octet scalars, quantifying the significance of the decay modes that appear in presence of the light gravitinos. Outside the restriction of low-scale SUSY breaking, in section \ref{fakesplitsusy} we describe how the decays of gluinos to gravitinos are modified in the limit of heavy squarks and when gauginos have both Majorana and Dirac masses, as in the context of Fake Split SUSY \cite{Dudas:2013gga,Benakli:2013msa}.

%%%%%%%%%%%%%%%
\section{Goldstino couplings with Dirac gauginos}
\label{SEC:GENERICCOUPLINGS}
%%%%%%%%%%%%%%%

In a supersymmetric gauge theory with Dirac gauginos, spontaneous breaking of supersymmetry induces couplings of the massive gravitino to all fields, including the new degrees of freedom that are required in order to attribute Dirac masses to the gauginos. According to the supersymmetric equivalence theorem \cite{cddfg}, in the high energy limit ($E\gg m_G$) the dynamics of the gravitino are dominated by the dynamics of its goldstino longitudinal component. In a low-energy effective Lagrangian with all superpartners integrated out, the goldstino couplings to SM fields consist of a set of universal, model-independent terms and the details of the microscopic theory account only for the determination of the coefficients of certain model-dependent terms, such as an effective dimension-8 coupling between two goldstinos and two fermions \cite{Brignole:1997pe}. If we are interested in collider signatures of the full theory, then we need to retain (some of) the superpartners, such as in the approach of nonlinear MSSM \cite{Brignole:2003cm,Komargodski:2009rz,Antoniadis:2010hs}.

The general expression of goldstino couplings can be obtained within the superfield formalism by use of a ``goldstino" superfield $X = \phi_X + \sqrt{2} \theta \psi_X + \theta \theta F_X$. The dynamics of the goldstino multiplet can be parametrised in terms of an effective Polonyi model:
\begin{align}
\C{L}_X =& \int d^4 \theta \bigg( 1 - \frac{m^2_X}{4f^2} X^\dagger X\bigg) X^\dagger X + \bigg( \int d^2 \theta f X + h.c. \bigg)\,.
\end{align}
The goldstino field $\psi_X$ will not be the true goldstino of the theory once we include couplings to matter, but it will ultimately mix with other fermions. Furthermore, in the limit of large sgoldstino mass $m_X$, we can integrate this scalar out which amounts to setting $\phi_X = \frac{\psi_X \psi_X}{2 F_X}$. This may not be necessary, but we should bear in mind that the dynamics of the scalar $\phi_X$ is equivalent to second order in the goldstino couplings. 

The Dirac gaugino mass operator in this language is given by
\be\label{Dmass}
\cL_{Dirac} = \frac{m_D }{\sqrt{2} f^2} \int d^4 \theta\, X^\dag D^\alpha X\, W_\alpha^a \mathbf{\Sigma}^a =  -\frac{m_D }{4\sqrt{2} f^2} \int d^2 \theta \,\ov{D}^2 D^\alpha (X^\dag X) \,W_\alpha^a \mathbf{\Sigma}^a \,,
\ee
where $W_\alpha^a = \lambda_\alpha^a + D^a \theta_\alpha + ...,$ is a gauge field strength with gauge index $a$ and $\mathbf{\Sigma}^a = \Sigma^a + \sqrt{2} \theta \chi_\Sigma^a + \theta \theta F_\Sigma^a $ an adjoint chiral superfield. In the case of an abelian gauge group, this will be a singlet superfield $\mathbf{S}(S,\chi_S,F_S)$. Defining $m_D = |m_D| e^{i\phi_D } $, $S_R \equiv \frac{1}{\sqrt{2}} (e^{i\phi_D}S + e^{-i\phi_D}\ov{S})$ and $S_I \equiv -\frac{i}{\sqrt{2}} (e^{i\phi_D}S - e^{-i\phi_D}\ov{S})$, the component expansion of eq. (\ref{Dmass}) up to second order in the goldstino delivers
\begin{align}\label{Dmasscomp}
\cL_{Dirac} =&\  2 |m_D| D S_R + \frac{|m_D|}{f^2} \bigg[-i D S_R( \partial_\mu \psi_X \sigma^\mu \ov{\psi}_X - \psi_X \sigma^\mu \partial_\mu \ov{\psi}_X ) \nn
\\
& +\frac{1}{2} S_I F_{\mu \nu}\epsilon^{\mu \nu \rho \lambda}  ( \psi_X  \sigma_\lambda \partial_\rho\ov{\psi}_X - \ov{\psi}_X \ov{\sigma}_\lambda \partial_\rho \psi_X ) +  S_R F_{\mu \nu} (\psi_X  \sigma^\mu \partial^\nu\ov{\psi}_X - \ov{\psi}_X  \ov{\sigma}^\mu \partial^\nu \psi_X)\bigg]\nn
\\
&+\bigg[\frac{m_D}{f^2} \bigg( - |F_X|^2 \lambda \chi  +  F^\dag_X \, \psi_X \lambda\,  F_S + S i  \lambda^\alpha \sigma^\mu_{\alpha \dot{\beta}} \partial_\mu ( F_X\ov{\psi}_X^{\dot{\beta}}) 
\nn
\\
&+ i S F^\dag_X \psi_X^\alpha \sigma_{\alpha \dot{\alpha}}^\mu D_\mu \ov{\lambda}^{\dot{\alpha}}  -\frac{1}{\sqrt{2}} \psi_X \chi\, F^\dag_X D + \frac{i}{2\sqrt{2}} F^\dag_X \psi_X^\alpha (\sigma^\mu \ov{\sigma}^\nu)_\alpha^\beta F_{\mu \nu} \chi_\beta\nn
\\
&+i(\chi  \sigma^\mu \partial_\mu \ov{\psi}_X) \,\psi_X \lambda - i\,\chi \partial_\mu \psi_X\,\lambda \sigma^\mu \ov{\psi}_X \bigg)+ h.c.\bigg]\,.
\end{align}
We observe that there are several differences with respect to the Majorana operator: 
\begin{enumerate}[(a)]
\item The sgoldstino does not enter at this order. Sgoldstino couplings always appear with at least one goldstino and in the limit $\phi_X \to \frac{\psi_X \psi_X}{2 F_X}$, this means that $\phi_X$ contributes at third order in goldstinos and beyond. 

\item A vacuum expectation value for the adjoint scalar $v_S \equiv \bra S_R \ket$ induces kinetic mixing between the goldstino and the gaugino 
\begin{align}\mathcal{L} \supset - \frac{ \sqrt{2}m_D v_S }{f} \bigg[ i \lambda \sigma^\mu \partial_\mu \ov{\psi}_X + i \psi_X \sigma^\mu D_\mu \ov{\lambda} \bigg] . \end{align} 
We discuss how this is removed by a supersymmetric field redefinition in the next subsection. 
\item The Dirac operator contains a coupling of two goldstinos to a gauge boson and the corresponding adjoint scalar. For a light enough scalar, this brings phenomenological signatures that are absent in the Majorana case. Furthermore, if this scalar acquires a vacuum expectation value, it apparently introduces a coupling between two goldstinos and a gauge boson which, by general considerations of goldstino Lagrangian constructions \cite{Brignole:1996fn,Ponton,Clark}, is expected to be absent.
\end{enumerate}

Dirac gauginos in the MSSM are accommodated by including three new adjoint chiral superfields: a singlet $S$ for the $U(1)$, a triplet $\overrightarrow{T}$ for the $SU(2)_L$ and an octet $\mathbf{O}$ for the $SU(3)_c$ gauge group. Then, apart from the Dirac mass operators of each gauge group, there can exist renormalisable couplings between the electroweak adjoints and the Higgs multiplets. The full effective Lagrangian that will be considered in this paper is given in appendix \ref{Leff}.

Before moving to section \ref{sgluondecays} and the study of the phenomenological implications of the new couplings of eq. (\ref{Dmass}), in the following subsection we clarify the apparent modification of the coupling of a goldstino to a Z-boson and a photon.

%%%%%%%%%%%%%%%
\subsection{Goldstino coupling to electroweak gauge bosons}
%%%%%%%%%%%%%%%

The vacuum expectation value of the adjoint scalar (in MSSM, the singlet and the neutral triplet) in the Dirac operator (\ref{Dmass}), induces an effective Fayet - Iliopoulos term:
\be
\cL_{FI}=-\frac{m_Dv_S }{8 f^2} \int d^2 \theta \,\ov{D}^2 D^\alpha (\ov{X} X) \,W_\alpha +h.c.\,.
\ee
This term contains a derivative coupling of two fermions $\psi_X$ to the field strength of a gauge boson,
\be\label{LVGG2}
\cL_{FI}\supset {m_{D}v_S\over f^2}\psi_X\sigma^{\nu}\ov{\psi}_X\partial^\mu F_{\mu\nu}\,,
\ee
which could apparently modify the coupling of the Z-boson or the photon to two goldstinos. Such a term is only permitted in the case of a theory with massive gauge bosons, and then the coupling should be of strength  \cite{Brignole:1996fn,Ponton,Clark,Komargodski:2009rz}:
\begin{align}
\mathcal{L} \supset&  \frac{\bra D \ket}{2f^2} m_A^2 \psi_X \sigma^\nu \ov{\psi}_X A_\nu \leftrightarrow  \frac{\bra D \ket}{2f^2} \psi_X \sigma^\nu \ov{\psi}_X \partial^\mu F_{\mu \nu}
\label{EQ:massiveDtermcoupling}\end{align}
which begs the question: now that the D-term is modified by the expectation value for $v_S$ and the Dirac mass term, 
\begin{align}
\bra D \ket = - 2 m_D v_S + ...\,,
\label{EQ:Dtermvev}\end{align}
 what is the D-term that should enter in (\ref{EQ:massiveDtermcoupling})? Since the low-energy effective theory of the gauge bosons, the Higgs fields and the goldstino should be independent of whether we have added Dirac mass terms for the gauginos or not, we expect that the coupling is not modified in the extended theory. Indeed, one way to illustrate this, is by performing a supersymmetric field redefinition. 
Let us consider the $U(1)$ gauge terms\footnote{For the case of $SU(2)$ we must perform a non-gauge-invariant shift.}:
\be\label{fil}
\cL=\cL_X+\left(\int d^2\theta\, {1\over 4}W^\alpha W_\alpha+{M\over 2f}XW^\alpha W_\alpha-{m^2\over 8f^2}\ov{D}^2D^\alpha (X^\dag X)W_\alpha +h.c.\right)
\ee
where $W_\alpha$ is an abelian field strength, and perform the following superfield redefinition
\begin{align}
V \rightarrow& V - \frac{m^2}{f^2} X^\dag X\,, \nn\\
W^\alpha\rightarrow& W^\alpha + {m^2\over 4f^2}\ov{D}^2D^\alpha (X^\dag X)\,.
\label{EQ:TRANSFORM}\end{align}
At the component level, this is
\bea
&&F_{\mu\nu}\rightarrow F_{\mu\nu}+{m^2\over f^2}\left[\partial_\mu(\ov{\psi}_X\ov{\sigma}_\nu\psi_X+2ix^\dag \partial_\nu x)-\partial_\nu(\ov{\psi}_X\ov{\sigma}_\mu\psi_X+2ix^\dag \partial_\mu x)\right]\,,\nn
\\
&&\lambda\rightarrow \lambda+{\sqrt{2}m^2\over f}\psi_X+{\sqrt{2}im^2\over f^2}(\ov{\psi}_X\ov{\sigma})^\alpha\partial_\mu x\,,\nn
\\
&&D\rightarrow D-2m^2+{im^2\over f^2}(\partial^\mu \psi_X\sigma_\mu \ov{\psi}_X- \psi_X\sigma_\mu \partial^\mu\ov{\psi}_X)-{2m^2\over f^2}|\partial_\mu x|^2\,.
\eea
The superpotential terms of the redefined Lagrangian are
\bea
\cL_W'&=&\int d^2\theta\, {1\over 4}W^\alpha W_\alpha+{M\over 2f}XW^\alpha W_\alpha+{Mm^2\over 4f^3}XW^\alpha\ov{D}^2D_\alpha (X^\dag X)\nonumber
\\
&&-{m^4\over 64f^4}\left[\ov{D}^2D_\alpha (X^\dag X)\right]^2+{Mm^4\over 32f^5}X\left[\ov{D}^2D_\alpha (X^\dag X)\right]^2+h.c.\,.
\eea
In the new basis, the gaugino and the goldstino have diagonal kinetic terms (in the original basis there is kinetic mixing), the shift by $2m^4$ in the vacuum energy is made explicit
%\footnote{Also in the soft scalar mass (by $m^2$), if we include a matter superfield charged under the abelian group.} 
(in the original basis it appears through the D-terms) and the coupling (\ref{LVGG2}) between two goldstinos and a gauge boson has been redefined away, and replaced with higher-derivative four-fermion couplings, while the D-term that appears in the coupling (\ref{EQ:massiveDtermcoupling}) does not contain a contribution (\ref{EQ:Dtermvev}) proportional to the Dirac gaugino mass.

We thus conclude that the coupling of a goldstino to electroweak gauge bosons is not modified in the extended theory where the gauginos acquire Dirac masses.

%%%%%%%%%%%%%%%
\section{Collider phenomenology}
\label{sgluondecays}
%%%%%%%%%%%%%%%

Since the mass of the gravitino is related to the SUSY breaking scale, phenomenological studies of low-scale SUSY breaking at colliders are based on searches for signals related to ultralight gravitinos. Gravitino production has been extensively studied in the past, following two main approaches:
\begin{enumerate}
\item \emph{Consider all supersymmetric particles to be heavy and integrate them out, leaving only the SM fields and the gravitino. Place limits on $\sqrt{f}$ from the resulting higher-dimensional operators. }
\item \emph{Make assumptions about the spectrum of superpartners and place limits on $\sqrt{f}$ as a function of their masses. }
\end{enumerate}
Clearly the first approach above is model-independent; for example \cite{Brignole:1998me} identified the effective higher-dimensional operators which generate monojet/monophoton signals and have universal coefficients. In this approach the nature of the supersymmetric spectrum is irrelevant: in particular, the results do not change whether we have Dirac or Majorana gauginos provided that the assumption that we can safely decouple the spectrum is justified. In this approach, the relevant process is direct gravitino pair production via an effective interaction of two partons and two gravitinos. The main signal is either monojet or monophoton from initial state radiation. However, the current bound of $\sqrt{f}> 240  $ GeV from monophoton events at LEP \cite{LEPmonophoton} is too low to justify this approach, when considered in conjunction with LHC search bounds for squarks and gluinos: it is somewhat perverse from a model-building perspective to assume that the superpartners are so much heavier than $\sqrt{f}$. We are then obliged to pass to the second approach, for which all studies so far have assumed Majorana gauginos; in this case current limits for light squarks give $\sqrt{f} \gtrsim 640 $ GeV \cite{ATLAS:2012zim}.

As we mentioned in the introduction, the gauginos of supersymmetric theories with a multi-TeV SUSY breaking scale are expected to acquire Dirac instead of Majorana masses. This implies that searches along the lines of \cite{ATLAS:2012zim} should be reconsidered. In this section we wish to highlight the main features of the novel phenomenology, while we leave the full phenomenological treatment for future work.

Among the new particles of the extended theory, the sgluon, the scalar octet partner of the gluino adjoint fermion, plays a central role. The mass of this particle is expected to be of the order the SUSY breaking scale\footnote{In explicit low-scale models the adjoint multiplet $\Sigma$ couples to the SUSY-breaking sector via a coupling  $W \supset \lambda_\Sigma \Sigma J_2$ \cite{Benakli:2008pg} where $\lambda_\Sigma $ should be large in order to generate large Dirac gaugino masses \cite{Gherghetta:2011na}; this then leads to the adjoint scalars having masses of order the scale of the SUSY-breaking dynamics if $\lambda_\Sigma$ takes its maximal value.} but in principle in can be even lighter than the gluino, as its current experimental bound is at 800 GeV \cite{Atlassgluon2}. It is the phenomenology of these particles that we would like to examine. In particular, we propose an additional option for light gravitino studies which is somewhere between options $1$ and $2$ above:

\indent $1.5.$\ \emph{Consider the effective theory of sgluons, gluinos and gravitinos, integrating out the squarks (except for perhaps the third generation).} 

The properties of the sgluons have been considered outside the context of low-scale SUSY breaking in \cite{Plehn,Choi:2008ub}. The main production process for these particles at the LHC is a tree level pair production via annihilation of two gluons, and indeed, the current experimental bound on its mass comes from searches for pair production where each sgluon decays to quark pairs. Much rarer single sgluon production can also occur, from pairs of gluons and quarks induced via a loop of squarks and gluinos. If kinematically allowed, the sgluon will decay predominantly via tree-level interactions to a pair of squarks or gluinos, and if not, the sgluon may decay to a pair of quarks or gluons via the same loop process mentioned above. Hence, pair production leads to signals with four or more jets, while single sgluon production has been typically assumed to lead to dijet production. 

For a low enough SUSY breaking scale, the presence of the almost massless gravitino can dramatically change  the phenomenology of a light sgluon. In particular, apart from the standard decay channels, the sgluon also decays to a pair of gravitinos and a gluon ($GGg$), leading to signals with fewer jets. Furthermore, if the lighter of the two sgluon mass eigenstates has a mass that is smaller than that of the squarks and gluinos, the tree-level decay to $GGg$ competes only with the loop-induced decay to two tops, thereby making the novel decay channel competitive even for a relatively high SUSY breaking scale.

Let us now compare the standard theory of Majorana gluinos with our extended theory of Dirac gluinos and sgluons. The effective theory of Majorana gluinos and gravitinos leads either to associated gravitino production, where one gravitino is produced in asssociation with one gluino which then decays to a gluon and a gravitino, or to indirect gravitino production, where two gluinos are produced and then decay to two gluons and two gravitinos. Since the gluino is the NLSP, the branching ratio of the tree level decay to gluon plus goldstino is close to 1. At leading order, the first process produces monojet events while the second produces dijet events. If initial and final state radiation are taken into account, both processes lead to multijet events with possibly one or two hard jets.

Once the effective theory is extended to that of Dirac gluinos, sgluons and gravitinos, new processes need to be considered in addition to the picture above. In particular, monojet events are produced not only by associated gravitino production but also by single sgluon production which subsequently decays to two gravitinos and a gluon. Dijet events are produced not only by gluino pair production but also by sgluon pair production, where again each sgluon decays to $GGg$. In the limit where the light sgluon is lighter than the gluino and the squarks, it decays mainly to $GGg$ at tree level and to two tops at one loop level. If the branching ratio of the light sgluon to $GGg$ is bigger than that to two top quarks, the effects of these new processes can be significant.

In the following we will study in detail the sgluon couplings and decay channels, compare their branching ratios and quantify their importance for monojet and multijet events.

%%%%%%%%%%%%%%%
\subsection{Conventional couplings and decays of a sgluon}
%%%%%%%%%%%%%%%

The mass terms of the scalar octet are given by
\bea
\cL_{M_O}&=&-m_O^2|O^a|^2-{1\over 2}(B_OO^aO^a+h.c.)-(m_{D}O^a+m_{D}^*O^{a*})^2\nn
\\
&=&-\tilde{m}_O^2|O^a|^2-{1\over 2}(\tilde{B}_OO^aO^a+h.c.)\,,
\eea
where $\tilde{m}^2_O=m^2_O+2|m_{D}|^2$ and $\tilde{B}_O=B_O+2m_{D}^2$. The two mass eigenstates are then
\be
O_{1}^a={1\over \sqrt{2}}\left(e^{{i\over 2}\phi_{\tilde{B}}}O^a+e^{-{i\over 2}\phi_{\tilde{B}}}O^{a*}\right)\,,\quad O_{2}^a=-{i\over \sqrt{2}}\left(e^{{i\over 2}\phi_{\tilde{B}}}O^a-e^{-{i\over 2}\phi_{\tilde{B}}}O^{a*}\right)\,,
\ee
with $M_{O_{1,2}}^2=\tilde{m}_O^2\pm |\tilde{B}_O|$.

In the following we present the decay rates of the two sgluons and we compare them with the novel decay to a pair of gravitinos and a gluon.

%%%%%%%%%
\subsubsection*{Sgluons decaying to squarks}
%%%%%%%%%

The sgluon couples to squarks via the D-term in (we take $m_{D}$ without loss of generality to be real -- otherwise we simply rotate the phase of $m_D$ into $\phi_O$)
\be
\cL\supset -\int d^2\theta {m_{D}\over 4\sqrt{2}f^2}\ov{D}^2D^\alpha(X^\dag X)W^a_\alpha \mathbf{O}^a\supset \sqrt{2}m_{D}(O^a+O^{a*})D^a_c\,,
\ee
which delivers
\begin{align}
\cL_{O^a\tilde{q}\tilde{q}}=& -2g_sm_{D}T^a_{xy} \sum_{\tilde{q}_L, \tilde{q}_R}(\tilde{q}_{Lxi}^*\tilde{q}_{Lyi}-\tilde{u}_{Rxi}^*\tilde{u}_{Ryi}-\tilde{d}_{Rxi}^*\tilde{d}_{Ryi})\left(\cos({\phi_{\tilde{B}}\over 2})O_1^a+\sin({\phi_{\tilde{B}}\over 2})O_2^a\right) \nn \\
=& -2g_sm_{D}T^a_{xy} \left(\cos({\phi_{\tilde{B}}\over 2})O_1^a+\sin({\phi_{\tilde{B}}\over 2})O_2^a\right) \\ 
&\times \sum_{i=1}^3 \sum_{j,k=1}^6 \bigg[ \bigg( Z_{ij}^* Z_{ik}   -  Z_{i+3\,j}^* Z_{i+3\,k} \bigg) \tilde{U}_{jx}^* \tilde{U}_{ky} + \bigg( W_{ij}^* W_{ik}   -  W_{i+3\,j}^* W_{i+3\,k} \bigg) \tilde{D}_{jx}^* \tilde{D}_{ky} \bigg] \nn
\end{align}
after switching to the mass eigenstate basis where
\begin{align}
\tilde{u}_{Li} =& Z_{ij} \tilde{U}_j, \qquad \tilde{u}_{Ri} = Z_{i+3\,j} \tilde{U}_j \,,\nn\\
\tilde{d}_{Li} =& W_{ij} \tilde{D}_j, \qquad \tilde{d}_{Ri} = W_{i+3\,j} \tilde{U}_j.
\end{align}
The decay rate is then given by
\bea
&&\Gamma_{O_1\to \tilde{U}_j\tilde{U}_k}= {\alpha_S m_{D}^2 \over M_{O_1}^2}\cos^2({\phi_{\tilde{B}}\over 2}) |p_{1jk}||Z^*_{ij}Z_{ik}-Z^*_{i+3j}Z_{i+3k}|^2
\\
&&\Gamma_{O_2\to \tilde{U}_j\tilde{U}_k}= {\alpha_S m_{D}^2 \over M_{O_2}^2} \sin^2({\phi_{\tilde{B}}\over 2})|p_{2jk}||Z^*_{ij}Z_{ik}-Z^*_{i+3j}Z_{i+3k}|^2\,,
\eea
where
\begin{align}
p_{ijk} \equiv& {1\over 2M_{O_i}}\sqrt{ (M_{O_i}^2 - m_{\tilde{U}_j}^2 - m_{\tilde{U}_k}^2)^2 - 4m_{\tilde{U}_j}^2m_{\tilde{U}_k}^2}  
\end{align}
and similarly for the down squarks. 
% \be
% \Gamma_{O_2\to \tilde{q}_i^*\tilde{q}_i}\!= {\alpha_S m_{D}^2 \over 2M_{O_2}}\sqrt{1-{4m_{\tilde{q}_i}^2\over M_{O_2}^2}}\sin^2({\phi_{O}\over 2})\,; \quad\Gamma_{O_1\to \tilde{q}_i\tilde{q}_i}\!= {\alpha_S m_{D}^2 \over 2M_{O_1}}\sqrt{1-{4m_{\tilde{q}_i}^2\over M_{O_1}^2}}\cos^2({\phi_{O}\over 2}).\nn
% \ee
When $\phi_{\tilde{B}}=0$ only $O_1$ can decay to squarks, since the coupling to $O_2$ vanishes.  % , which we will assume in the following, only the heavier of the two sgluons can decay to squarks.

\subsubsection*{Sgluons decaying to gluons}

Although there is no tree-level coupling of the sgluons to gluons, loops involving the squarks via the D-term coupling yield
\begin{align}
\mathcal{L} \supset & \frac{2g_s^3}{16\pi^2} \bigg[  \frac{m_D}{M_{O_1}^2} \lambda_{g_1} \cos (\frac{\phi_{\tilde{B}}}{2}) O^a_1 + \frac{m_D}{M_{O_2}^2}  \lambda_{g_2} \sin (\frac{\phi_{\tilde{B}}}{2}) O^a_2\bigg] d^{abc} F_{\mu\nu}^b F^{c\, \mu\nu} .
\label{EQ:gluoncoupling}
\end{align}
where
\begin{align}
%\lambda_{g_i} \equiv& 2  \bigg(\tau_{Li} f(\tau_{Li}) - \tau_{Ri} f(\tau_{Ri})\bigg)
\lambda_{g_i} \equiv & \sum_{j=1}^3 \sum_{k=1}^6 \bigg(Z_{jk}^* Z_{jk}    -  Z_{j+3\,k}^* Z_{j+3\,k} \bigg) (\tau^U_{ik} f( \tau_{ik}^U) -1) \nn\\
 + &\sum_{j=1}^3 \sum_{k=1}^6 \bigg(W_{jk}^* W_{jk}    -  W_{j+3\,k}^* W_{j+3\,k} \bigg)( \tau^D_{ik} f( \tau_{ik}^D)  -1)
\end{align}
and
\begin{align}
\tau^U_{ik} \equiv& \frac{4 m_{\tilde{U}_k}^2}{M_{O_i}^2},\qquad \tau^D_{ik} \equiv \frac{4 m_{\tilde{D}_k}^2}{M_{O_i}^2} \nn\\
f(\tau) \equiv& \left\{ \begin{array}{cl} (\sin^{-1} (1/\sqrt{\tau}))^2 & \tau \geq 1 \\ -\frac{1}{4} \bigg[ \ln \frac{1+\sqrt{1-\tau}}{1-\sqrt{1-\tau}} - i\pi \bigg]^2 & \tau < 1 \end{array} \right.\,.
\end{align}
These expressions generalise those of \cite{Plehn,Choi:2008ub}; in particular, they treat the sgluons as two sets of real scalars rather than one complex scalar. 

% The sgluon can decay to gluons only via squark loops, therefore this channel is open only for $O_1$ (the heavier mode). The decay rate in the absence of L-R mixing is given by \cite{Choi:2008ub}
% \be
% \Gamma_{O_1 \rightarrow gg} \ = \ \frac{5 \alpha_s^3}{192 \pi^2} \frac{m_{D}^2}{M_{O_1}} \bigg|\sum_{\textrm{squarks}} [ \tau_L f(\tau_L) - \tau_R f(\tau_R) ] \bigg|^2 \nn
% \ee
% where $\tau_{L,R} \equiv 4 m_{\tilde{q}_{L,R}}^2/M_{O_1}^2$ and
% \begin{align}
% f(\tau) \equiv& \left\{ \begin{array}{cl} (\sin^{-1} (1/\sqrt{\tau}))^2 & \tau \geq 1 \\ -\frac{1}{4} \bigg[ \ln \frac{1+\sqrt{1-\tau}}{1-\sqrt{1-\tau}} - i\pi \bigg]^2 & \tau < 1 \end{array} \right.\,.
% \end{align}
If we neglect mixing between `left' and `right' squarks we can write 
\begin{align}
\lambda_{g_i} = & \sum_{\tilde{q}}  \bigg( \tau_{i \tilde{q}_L}f( \tau_{i \tilde{q}_L}) - \tau_{i\tilde{q}_R}f( \tau_{i\tilde{q}_R}) \bigg).
\end{align}
We observe that in the limit of degenerate masses for the `left' and `right' squarks, the decay rate diminishes. Finally, the widths for the sgluons are given by
\begin{align}
\Gamma (O_1 \rightarrow gg) =& \frac{5 \alpha_s^3}{192 \pi^2} \frac{m_{D3}^2}{M_{O_1}}  \cos^2 (\frac{\phi_{\tilde{B}}}{2})|\lambda_{g_1}|^2, \quad \Gamma (O_2 \rightarrow gg) = \frac{5 \alpha_s^3}{192 \pi^2} \frac{m_{D3}^2}{M_{O_2}}  \sin^2 (\frac{\phi_{\tilde{B}}}{2})|\lambda_{g_2}|^2.\nn
\end{align}

\subsubsection*{Sgluons decaying to tops}

The final conventional coupling of the sgluons is that to quark-antiquark pairs, which is again generated at one loop. However, these decays are suppressed by the quark masses, so the only substantial decay channel is to top quarks. In this case, for simplicity we shall neglect left-right squark mixing (a very good approximation in Dirac gaugino models, and exact in models preserving R-symmetry) and take $\phi_O = 0$ (i.e. we shall assume no CP violation in the sgluon-gluino sector or that both the $B_O$ and $m_D$ are generated by the same process, such as in gauge-mediation). 

There are two triangle loop integrals that contribute to this coupling, one with two gluinos and one squark and the other with two squarks and one gluino. The sgluon $O_1$ receives contribution from both diagrams, while $O_2$ can decay only via the 2 gluinos - 1 squark loop. The contribution of the loop diagram is proportional to the mass of the two quarks, so top quarks receive the biggest contribution. We can parametrise the couplings in the effective Lagrangian as
\begin{align}
\mathcal{L} \supset &  c_{1\ov{t}t} \bar{t} O_1 t + c_{2\ov{t}t} i \bar{t} O_2\gamma_5 t.
\end{align}
Then the decay rate of $O_i$ to two top quarks is
\begin{align}
\Gamma (O_1 \rightarrow t\ov{t}) =& \frac{|c_{1\ov{t}t} |^2}{16\pi}M_{O_1}  \left(1-{4m_t^2\over M_{O_1}^2}\right)^{3\over 2}, \quad \Gamma (O_2 \rightarrow t\ov{t}) = \frac{|c_{2\ov{t}t} |^2}{16\pi} M_{O_2} \sqrt{1-{4m_t^2\over M_{O_2}^2}}
\end{align}
We find, similar to \cite{Choi:2008ub,Plehn} that
\begin{align}
 c_{1\ov{t}t} =& \frac{3 g_s^3}{16\pi^2}  m_D m_t I_S,\qquad c_{2\ov{t}t} =  \frac{3 g_s^3}{16\pi^2}  m_D m_t I_P
\end{align}
where 
\be
I_P=V_{3i}^LC_{0L}V_{i3}^{L\dag}-V_{3i}^RC_{0R}V_{i3}^{R\dag}.
\ee
The decay rate of $O_2$ to two top quarks is thus
\be\label{O2tt}
\Gamma_{O_2\to\ov{t}t}={9\alpha_s^3\over 64\pi^2}m_{D}^2m_t^2M_{O_2}\sqrt{1-{4m_t^2\over M_{O_2}^2}}I_P^2
\ee
with $V_{3i}^{L,R}=U^{u_{L,R}\,\dag}_{3j}Z^{L,R}_{ji}$ being the flavour rotation matrix ($U_{ij}$ is the quark and $Z_{ij}$ the squark diagonalizing matrix) and $C_{0L,R}[m_t^2,M_{O_2}^2,m_t^2;m^2_{\tilde{u}_i^{L,R}},m^2_{D},m^2_{D}]$ the usual P.-V. functions. We have 
\be
I_S=V_{3i}^LI_S^LV_{i3}^{L\dag}-V_{3i}^RI_S^RV_{i3}^{R\dag}
\ee
with
\begin{align}
I_S^{L,R}=&{1\over 9(M_{O_1}^2-4m_t^2)}\times\Big[(2m_{D}^2+2m_t^2-2m_{\tilde{u}_i^{L,R}}^2)C_{0}(m_t^2,M_{O_1}^2,m_t^2;m_{D}^2,m_{\tilde{u}_i^{L,R}}^2,m_{\tilde{u}_i^{L,R}}^2) \nn
\\
&(18m_{D}^2-9M_{O_1}^2+18m_t^2-18m_{\tilde{u}_i^{L,R}}^2)C_{0}(m_t^2,M_{O_1}^2,m_t^2;m_{\tilde{u}_i^{L,R}}^2,m_{D}^2,m_{D}^2)
\\
&+18B_0(m_t^2;m_{\tilde{u}_i^{L,R}}^2,m_{D}^2)-2B_0(m_t^2;m_{D}^2,m_{\tilde{u}_i^{L,R}}^2) -18B_0(M_{O_1}^2;m_{D}^2,m_{D}^2)\nn
\\
&+2B_0(M_{O_1}^2;m_{\tilde{u}_i^{L,R}}^2,m_{\tilde{u}_i^{L,R}}^2)\Big]\,. \nn
\end{align}
Here we are using the following definitions of the Pessarino-Veltman functions:
\begin{align}
B_0 (p^2,m_1^2, m_2^2) \equiv&-16\pi^2i\int {d^dq\over (2\pi)^{4-2\epsilon}}\frac{1}{(q^2 - m_1^2)}\frac{1}{(q-p)^2 - m_2^2} \\
C_0 (k_1^2, (k_1 - k_2)^2, k_2^2; m_1^2, m_2^2, m_3^2) \equiv & \frac{1}{i\pi^2} \int d^4 q \frac{1}{q^2 - m_1^2} \frac{1}{(q+ k_1)^2 - m_2^2} \frac{1}{(q+ k_2)^2 - m_3^2} \nn
\end{align}
where our definition for $C_0$ differs by permutation of the momenta and masses compared to those used in the comparable expressions in \cite{Plehn}; the above agrees with both that reference and \cite{Choi:2008ub}.

Hence the decay of $O_1$ is
\be
\Gamma_{O_1\to\ov{t}t}={9\alpha_s^3\over 64\pi^2}{m_{D}^2m_t^2\over M_{O_1}}\sqrt{1-{4m_t^2\over M_{O_1}^2}}(M_{O_1}^2-4m_t^2)I_S^2 .
\ee

Just as in the decay to gluons, this both decay rates vanish in the limit of degenerate `left' and `right' squarks.

%%%%%%%%%%%%%%%
\subsection{Novel sgluon decays in the effective theory of sgluons, gluinos and gravitinos}
\label{SEC:novel}
%%%%%%%%%%%%%%%

As we have mentioned above, if the scale of SUSY breaking lies in the multiTeV range, the presence of a light gravitino opens up the novel decay of a sgluon to two gravitinos and a gluon ($O\to GGg$). This process is described by the effective $OGGg$ coupling that is contained in the Dirac mass operator (\ref{Dmass}) and via the intermediate production of a gluino and a gravitino ($O\to G\tilde{g}\to GGg$). The explicit expression of the effective coupling is
\begin{align}
\mathcal{L}_{OGGg}=\frac{m_{D}}{f} \partial^\mu(G \sigma^\nu \ov{G}) G^a_{\mu\nu} O^a_1 + \frac{ m_{D}}{2f} \epsilon^{\mu \nu \rho \lambda} \partial_\rho (G \sigma_\lambda \ov{G}) G^a_{\mu\nu} O^a_2\,,
\label{EQ:OctetOperators}
\end{align}
where $G$ is the true goldstino and $G_{\mu\nu}^a$ is the gluon field strength. The relevant terms of the second process ($O\to G\tilde{g}\to GGg$) involving a purely Dirac gluino ($M_O=M_3=0$) can be extracted from appendix \ref{Leff}:
\begin{align}\label{EQ:OctetOperators2}
\mathcal{L}
\supset&\left( \frac{m_D}{\sqrt{2} f} G \sigma^{\mu\nu} \chi^a G^a_{\mu\nu} + \frac{i}{\sqrt{2}f} M_{O_2}^2  O_2^a  G \chi^a  - \frac{1}{\sqrt{2}f} ( M_{O_1}^2 - 2 m_D^2) O_1^a  G \chi^a + h.c.\right) \nn\\
& +\frac{ i m_D}{\sqrt{2} f} \partial_\mu O_1 ( G \sigma^\mu \ov{\lambda} - \lambda \sigma^\mu \ov{G})  +  \frac{ m_D}{\sqrt{2} f} \partial_\mu O_2 ( \lambda \sigma^\mu \ov{G} + G \sigma^\mu \ov{\lambda} ) \, .
\end{align}
If the octet scalar is heavier than the gluino, then it will decay to gluino plus gravitino with width
\begin{align}
\Gamma(O_i \rightarrow \tilde{g} G) =& \frac{(M_{O_i}^2 - m_{D}^2)^4}{32 \pi f^2 M_{O_i}^3} \,.
\end{align}

With the current bounds on the gluino mass this means looking for rather heavy sgluons, which will have rather low production rates. Instead we shall consider the lower mass regime, where the sgluon is lighter than the gluino. If we integrate out the gluino, we find that the two contributions from eq. (\ref{EQ:OctetOperators}) and (\ref{EQ:OctetOperators2}) cancel each other out at leading order in the momentum (the leading operator then being a higher-derivative one) which could be seen by starting from a non-linear scalar superfield. The decay width is then
%\footnote{$p_{1,2}$ are the momenta of the goldstinos and $p_3$ that of the gluon.}:
%\be
%\Gamma_{O_i}\!=\!-{m_{D3}^2\over  32\pi^3f^4}\!\!  \int^{M_{O_i}/2}_{0}\!\!\!\!\!\!\!\!\!\!\!dE_1\!\!\int^{-1}_{+1}\!\!\!\!\!d\!\cos\theta {E_1(M_{O_i}-2E_1) \over (M_{O_i}+E_1(\cos\theta-1))^2}p_1\cdot p_2\big( (p_1\cdot p_3)^2+(p_2\cdot p_3)^2 \big)
%\ee
%which simplifies to
%This allows for the three body decay rate of each sgluon eigenstate to two goldstinos and a gluon
\be
\Gamma (O_i\to GGg)={m_{D}^2 M_{O_i}^7\over 15360\pi^3f^4} g(y) \,,
\ee
where $y  \equiv \frac{M_{O_i}^2}{m_{D}^2} $ and
\begin{align}
g(y) \equiv&\ \frac{60 \left(3-y\right) \left(1-y\right)^3 \log \left(1-y\right)}{y^{5}} +\frac{6 y^4-155 y^3+480 y^2-510 y+180}{y^{4}} \nn\\
=&\  \frac{2}{7} y^2 + \frac{3}{14} y^3 + \frac{1}{7} y^4 + ... \nn\\
g(1) =&\ 1\,.
\end{align}
The decay rate is suppressed by eight powers of the SUSY breaking scale and thus it might be expected to be small. However, as we saw above, if the light sgluon is lighter than the gluino, the decay to $GGg$ competes only with a one-loop decay to a pair of quarks. Furthermore, this one-loop suppressed decay diminishes in the limit of heavy gluinos, heavy squarks and degenerate left-right squarks.

\begin{figure}
\includegraphics[scale=0.5]{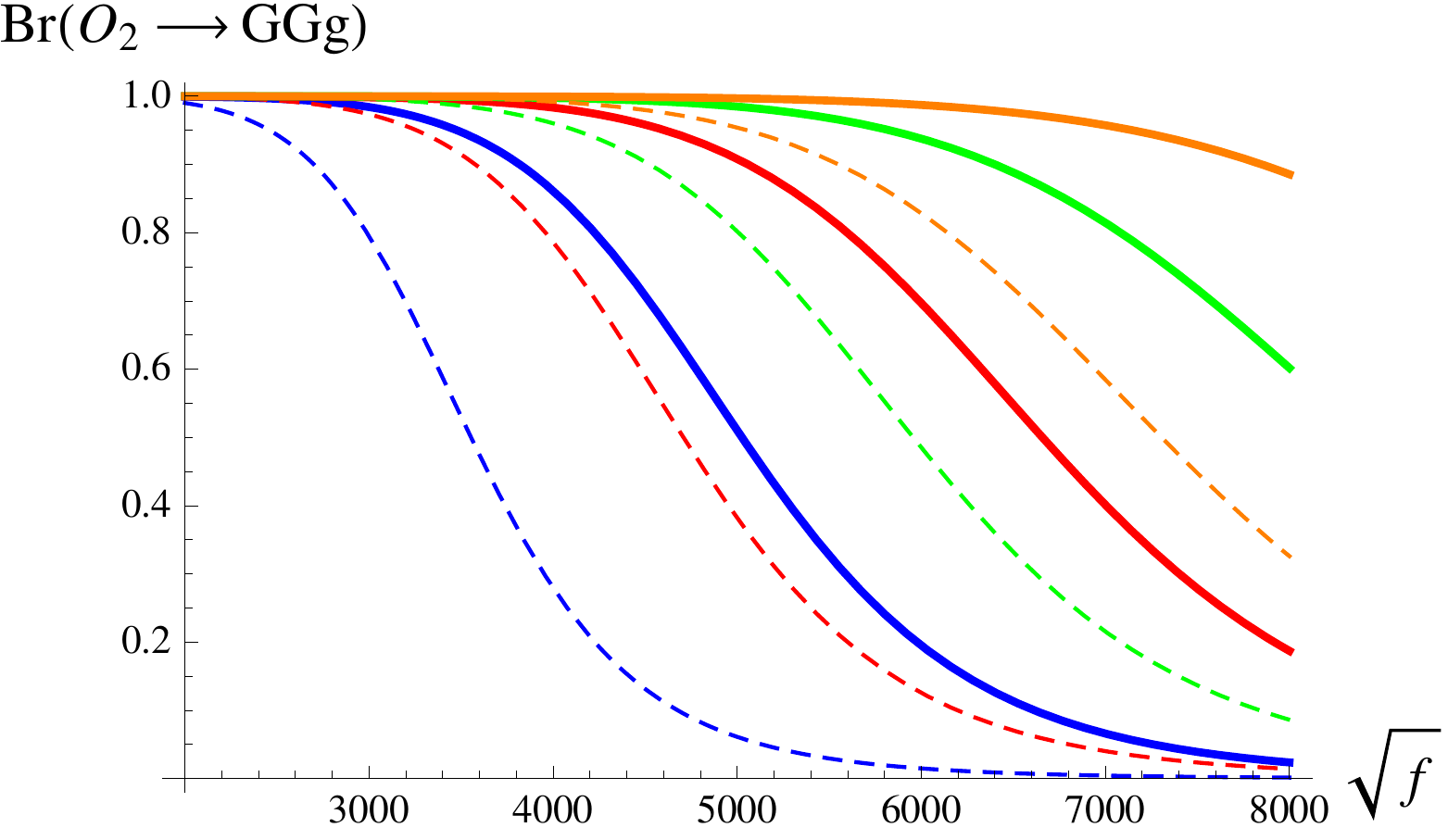}
\includegraphics[scale=0.5]{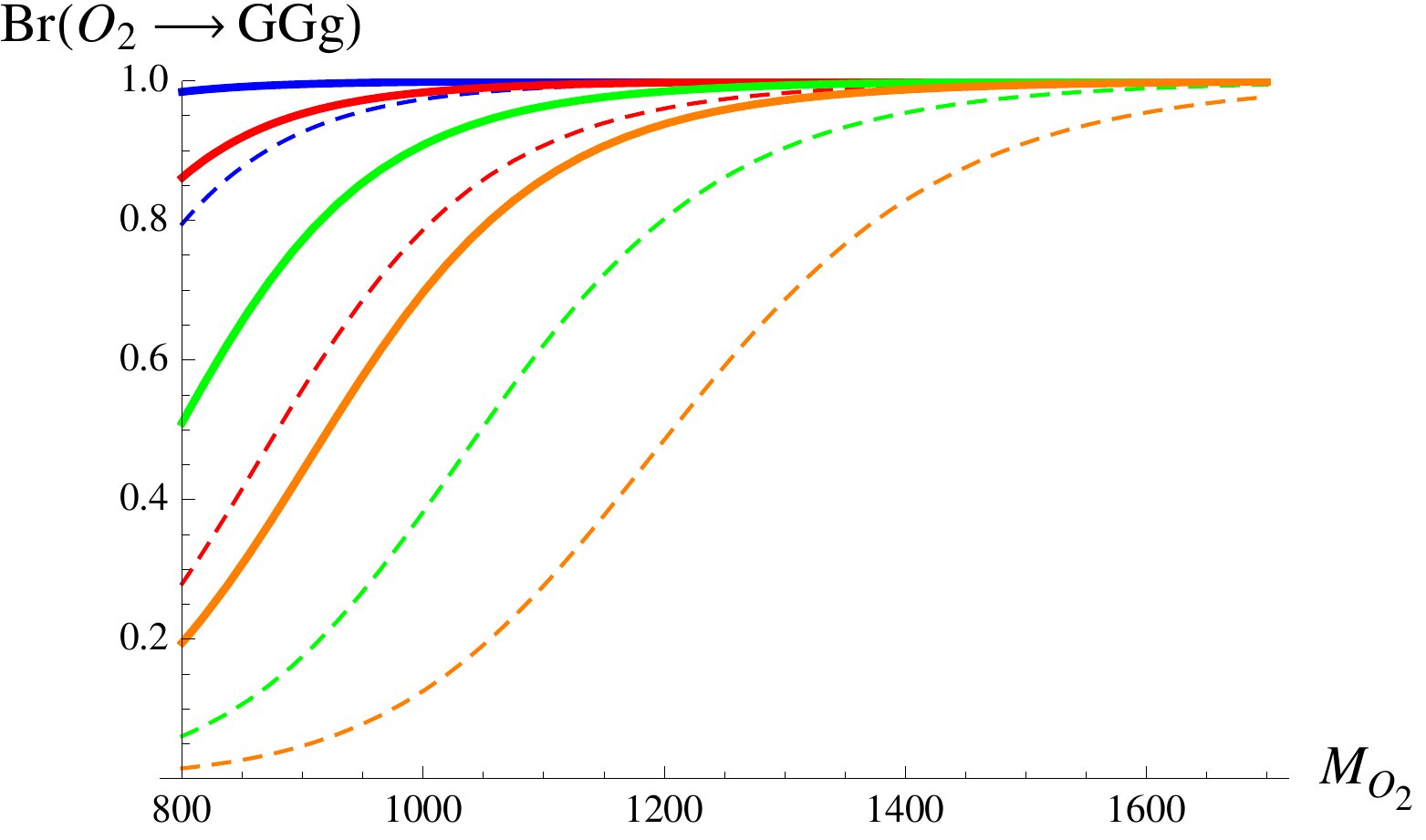}
\caption{The dependence of the branching ratio of the decay of a light sgluon to two gravitinos and a gluon on the SUSY breaking scale (left plot) and the mass of the light sgluon (right plot). Full lines are drawn for $\delta \tilde{m}=M_Z=90\,$GeV and dotted lines for $\delta \tilde{m}=180\,$GeV. Also, we have taken $m_{\tilde{g}}=1.7\,$TeV and $m_{\tilde{q}}=1.5\,$TeV. Left plot: (Blue, Red, Green, Orange): $M_{O_2}=(0.8\,$TeV, $1\,$TeV, $1.2\,$TeV, $1.4\,$TeV). Right plot: (Blue, Red, Green, Orange): $\sqrt{f}=(3\,$TeV, $4\,$TeV, $5\,$TeV, $6\,$TeV).}
\label{FIG:BrOtoGGg}
\end{figure}

In the limit $m_{D},m_{\tilde{q}_L},m_{\tilde{q}_R}\gg m_t,M_{O_2}$, we can obtain an analytic comparison of the two decay channels. Neglecting flavour mixing between squarks, we obtain the approximation (for $m_{\tilde{t}_{L,R}}\neq m_{D}$)
\be
I_P = \frac{ m_{D}^2 - m_{\tilde{t}_R}^2 - m_{\tilde{t}_R}^2 \ln( {m_{D}^2\over m_{\tilde{t}_R}^2})}{(m_{D}^2 - m_{\tilde{t}_R}^2)^2}-(R\to L)\,.
\ee
Taking $m_{\tilde{t}}^2=m_{\tilde{t}_L}^2=m_{\tilde{t}_R}^2-\delta\tilde{m}^2$, the above expression at first order in ${\delta\tilde{m}^2\over m_{\tilde{t}}^2}$ becomes
\be
I_P \simeq {\delta\tilde{m}^2\over m_{\tilde{t}}^4}{(1+x)(-1+x-\ln x)\over (x-1)^3}\,,
\ee
where $x=m_{D}^2/m_{\tilde{t}}^2$. The ratio of the decay rates in this limit is
\be
\frac{\Gamma_{O_2\to tt}}{\Gamma_{O_2\to GGg}}=7560\pi\alpha_S^3 \frac{m_t^2\delta\tilde{m}^4f^4}{ y^2 m_{\tilde{t}}^8M_{O_2}^6}\sqrt{1-{4m_t^2\over M_{O_2}^2}}{(-1+x^2-(1+x)\ln x)^2 \over (x-1)^6}\,.
\ee
In figure \ref{FIG:BrOtoGGg} (now without approximations, and loop functions evaluated using {\tt LoopTools} \cite{Hahn:1998yk}) we observe that, depending on the splitting between the squark masses, we can have large Br$(O_2\to GGg)$ even for a relatively high SUSY breaking scale. For example, if we make the assumption that the squark masses are split only by the electroweak contributions $\delta\tilde{m}^2\simeq M_Z^2$, then Br$(O_2\to GGg) >$ Br$(O_2\to tt)$ for a $1\,$TeV sgluon for a SUSY breaking scale as high as $\simeq$ 6.5 TeV. Also, as long as the gluinos and the squarks are heavier than the sgluon, changing their mass barely affects the above results.

%%%%%%%%%%%%%%%
\subsection{Monojet events}
\label{SEC:Monojet}
%%%%%%%%%%%%%%%

The above analysis suggests that in the interesting scenario of a sgluon that is lighter than the squarks and the gluinos, its decay to two gravitinos and a gluon will dominate over the conventional decay to two top quarks, even for a relatively high SUSY-breaking scale. As a consequence, both the phenomenology of a light sgluon and the phenomenology of light gravitino production will be modified.

Phenomenological studies of a light sgluon have mainly focused on signals with two jets and tops (from single sgluon production) and four or more jets and tops (from sgluon pair production). If the SUSY breaking scale is low enough then a light sgluon will produce predominantly monojets (from single sgluon production) and di/trijets (from sgluon pair production), drastically modifying the sgluon phenomenology.

If these processes are comparable to the conventional ones of gravitino production, the phenomenology of a light gravitino will also be modified. In particular, monojet events will be produced both by gravitino/gluino associated production and by single sgluon production. Also, dijet events will be produced both by gravitino pair production and by sgluon pair production.

\begin{figure}
\includegraphics[width=\textwidth]{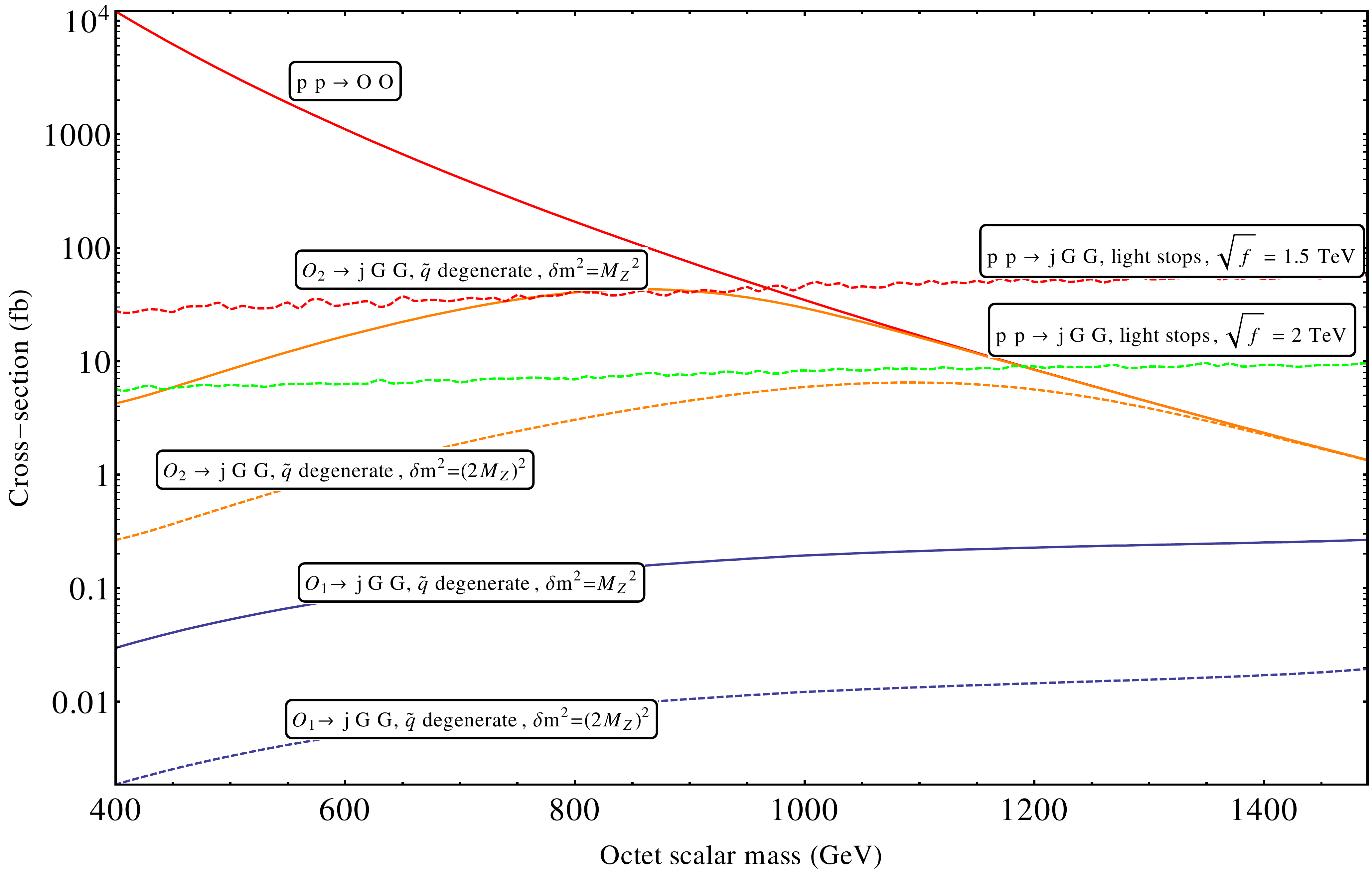}
\caption{Cross-sections at LHC13 for goldstino events when one octet scalar is light as a function of the octet scalar mass, with the total double octet production cross-section given as reference (labelled $p\ p \rightarrow O\ O)$. Events where two sgluons are produced and \emph{at least one}  decays to goldstinos (as opposed to two jets) are labelled $O_1 \rightarrow j G G$ and $O_2 \rightarrow j G G$. $\sqrt{f} = 7.5$ TeV was chosen since then $m_{\tilde{q}} \sim m_{\tilde{g}} \sim 0.2 \sqrt{f} \sim 1.5 $ TeV, with the squark masses  varying from a common SUSY-breaking mass as $\sqrt{m_{\tilde{q}}^2 \pm \frac{1}{2} M_Z^2},\sqrt{m_{\tilde{q}}^2 \pm 2 M_Z^2} $ as discussed the text (i.e. $\delta m^2 = M_Z^2, (2M_Z)^2$). Monojet events are labelled $p\ p \rightarrow j\ G\ G$; for these, two different, lower, values of $\sqrt{f}$ are shown, and the spectrum of other sparticles has the first two generations of squarks and  the right-handed squarks of the third generation at $2$ TeV, with left-handed third-generation squarks at $755$ GeV. In all cases the gluino mass was fixed at $1500$ GeV.}
\label{FIG:xsections}
\end{figure}

To illustrate the possibility of observing gravitino events from the octet scalars at the LHC, we implemented the model in {\tt Feynrules} \cite{Degrande:2011ua,Alloul:2013bka} and {\tt MadGraph} \cite{Alwall:2014hca}. We also used the {\tt CalcHEP} \cite{Belyaev:2012qa} output from Feynrules to verify analytical expressions for the widths. We then plotted the cross-sections for monojet -- only involving the gluon couplings, not the quark couplings  which do not interfere\footnote{Events involving the quark couplings to goldstini do not interfere with the events via the gluon couplings, and also do not change compared to the Majorana case -- we therefore excluded these events in order to show the contrast.} -- plus gravitinos production from proton-proton collisions at 13 TeV for several scenarios in figure \ref{FIG:xsections}. Note that in this scenario it is impossible to separate the gluino-mediated events from the sgluon ones (there are many diagrams in the process which do not involve the octet scalar at all). Hence we also require the (Dirac) gluino partial widths, which can be rather large:
\begin{align}
\Gamma (\tilde{g} \rightarrow g G ) =& \frac{m_{D}^5}{32 \pi f^2}\nn\\
\Gamma (\tilde{g} \rightarrow O_i G ) =&  \frac{(m_{D}^3 - m_{O_i}^2)^4}{64 \pi m_{D3}^3 f^2}\nn\\
\Gamma (\tilde{g} \rightarrow \tilde{q}_i^* q_i / \tilde{q}_i \ov{q}_i) =& \frac{\alpha_s}{2}  \frac{(m_{D}^2 + m_{q_i}^2 - m_{\tilde{q}_i}^2) \sqrt{(m_{D}^2 - m_{q_i}^2 - m_{\tilde{q}_i}^2)^2 - 4  m_{q_i}^2m_{\tilde{q}_i}^2}}{4 m_{D}^3}.
\end{align}
In the final line, we have shown the decays of a gluino to a quark/squark of one generation and chirality. 

From figure \ref{FIG:xsections} we conclude that the bounds on the sgluon $O_2$ in particular may be dramatically weakend in scenarios of low-scale supersymmetry breaking, since the branching ratio to a gravitino may be large. Since the mass of this state (assuming CP conservation) is $m_O^2 - B_O$ it is easily conceivable that this state is light if the conventional and $B$-type masses are of similar magnitude. However, this conclusion is strongly dependent on the mass-splitting between the squarks. On the other hand, if there is large splitting between the squarks, in particular if stops are light, then the coupling of $O_1$ to gluons from equation (\ref{EQ:gluoncoupling}) can be large. In this case, a significant number of monojet events could be observed. We see from the figure that \emph{increasing} the mass of $O_1$ toward the mass of the gluino increases the number of events by a factor of a few. One point to note is that for the same parameters (gluino mass, squark masses) we obtain cross-sections for monojets in the \emph{Majorana} gaugino case of $44\mathrm{fb}$ and $282\mathrm{fb}$ for $\sqrt{f} = 2,1.5$ TeV respectively, which are substantially larger than those we observe here -- indicating that in the Dirac case, as we expect, the bounds are weaker. However, since the sgluon is light the kinematics of the events should be different in the Dirac case which could potentially aid discovery; we leave investigation of this to future work.

%%%%%%%%%%%%%
\section{Gluino decay to gravitinos and Fake Split SUSY}
\label{fakesplitsusy}
%%%%%%%%%%%%%

So far we have discussed the phenomenology of Dirac gluinos, sgluons and gravitinos within the context of low-scale SUSY breaking. However, the interaction between gluinos and gravitinos has important consequences in more conventional models of SUSY breaking, too. In particular, if the gluino is heavier than the gravitino, it can decay directly to gravitinos and quarks or gluons as an alternative to other supersymmetric decays. The corresponding branching ratio can be significant in split-type models where the squarks are much heavier than the gluinos. Furthermore, in a general treatment of gaugino masses, such as in Fake Split SUSY \cite{Dudas:2013gga,Benakli:2013msa}, gluinos can have both a Majorana and a Dirac type mass while the coupling to the gravitino will be modified with respect to the pure Majorana case.

In particular, after rotating from the basis $(\lambda_3^a,\chi_O^a)$ to the mass eigenstate basis $\lambda_3 = R_{11} \tilde{g}_1 + R_{12} \tilde{g}_2$, $\chi_O = R_{21} \tilde{g}_1 + R_{22} \tilde{g}_2 $, the coupling of the gluinos to a gluon and a goldstino becomes
\be
\frac{1}{\sqrt{2}f} ( M_3 \lambda^a_3  + m_{D3} \chi_O^a)  \sigma^{\mu \nu} G G_{\mu\nu}^a+h.c.\to R_{11} \frac{m_{\tilde{g}_1}}{\sqrt{2}f} \tilde{g}_1^a  \sigma^{\mu \nu} G G_{\mu\nu}^a +  R_{12}\frac{m_{\tilde{g}_2}}{\sqrt{2}f} \tilde{g}_2^a  \sigma^{\mu \nu} G G_{\mu\nu}^a +h.c.\nn
\ee
where the two gluino masses $m_{\tilde{g}_i}$ are expressed in terms of $(M_O,m_{D3},M_3)$ and $M_O$ is the adjoint fermion mass, see eq. (\ref{Leff9}). The effective coupling of the gluinos to two quarks and a goldstino is obtained after integrating out the heavy squarks in the following terms
\begin{equation}
\cL \supset - {m_{\tilde{q}}^2\over f} \tilde{q}^* q G - \sqrt{2} g_s \tilde{q}^* q \lambda = - {m_{\tilde{q}}^2\over f} \tilde{q}^* q G - \sqrt{2} g_s R_{11}\tilde{q}^* q \tilde{g}_1 - \sqrt{2} g_s R_{12}\tilde{q}^* q \tilde{g}_2 
\end{equation}
to obtain
\begin{align}
\mathcal{L} \supset&  -\frac{\sqrt{2}g_s}{f}  R_{11} \ov{q} \ov{G}\,q \tilde{g}_1 -\frac{\sqrt{2}g_s}{f}  R_{12} \ov{q} \ov{G}\,q \tilde{g}_2+ h.c. 
\end{align}
In fact, even if we include left-right mixing among the squarks (which, unlike in Split SUSY, may be significant in Fake Split SUSY) the result is that in the leading order in $f^{-1}$, the goldstino couplings are given by, in Dirac notation now where $\Lambda^a_k \equiv \left(\begin{array}{c} \tilde{g}_k^a \\ \ov{\tilde{g}}_k^a \end{array}\right) $,
\begin{align}
%\mathcal{L} \supset& -\frac{\sqrt{2} g_s }{f} T^a_{ij} \bigg[ R_{1k} ( \ov{q}_i P_R G)(\ov{\Lambda}^a_k P_L q_j) - R_{1k}^* ( \ov{q}_i P_L G)(\ov{\Lambda}^a_k P_R q_j) + h.c. \bigg]
\mathcal{L} \supset& -\frac{\sqrt{2} g_s }{f} T^a_{ij} \bigg[ R_{1k} ( \ov{q}_{Li} G)(\ov{\Lambda}^a_k q_{Lj}) - R_{1k}^* ( \ov{q}_{Ri}  G)(\ov{\Lambda}^a_k  q_{Rj}) + h.c. \bigg]
\end{align}
i.e. the decays to left- and right-handed components of the quarks are independent. 

% \bea
% \cL &\supset& -\sqrt{2}g_s[ \tilde{q}_{Lxi}^*T^a_{xy}( R_{11}\psi_1^{a\alpha}\!+\!R_{12}\psi_2^{a\alpha} )  q_{Lyi\alpha} -\tilde{q}_{Rxi}T_{xy}^{a*}( R_{11}\psi_1^{a\alpha}\!+\!R_{12}\psi_2^{a\alpha} )q_{Ryi\alpha}^c]\nn
% \\
% &&- {m_{\tilde{q}_L}^2\over f} \tilde{q}_L q_LG -{ m_{\tilde{q}_R}^2\over f} \tilde{q}_R q_RG+h.c.\nn
% \eea

We can now consider the decays of the gluino in different scenarios. The general expressions of the decay widths to gluons and to a flavour $j$ of light quarks (i.e. ignoring the top, which will have a further suppressed width due to reduced phase space) is given by
\begin{align}
\Gamma (\tilde{g}_i \rightarrow g G) =& |R_{1i}|^2 \frac{m_{\tilde{g}_i}^5}{16 \pi f^2}\,, \nn\\
\Gamma (\tilde{g}_i \rightarrow q_j \ov{q}_j G) =& |R_{1i}|^2\frac{g_s^2 m_{\tilde{g}_i}^5}{1536 \pi^3 f^2},
\end{align}
where we have summed over the two chiralities of the quark $j$. 
In the case of a purely Majorana gluino where $R_{12} = 0, R_{11} = 1$, we recover the standard expressions \cite{Gambino:2005eh}. For a purely Dirac gluino, $R_{11} = \frac{1}{\sqrt{2}}, R_{12} = -\frac{i}{\sqrt{2}}$ and since the gluino can be thought of as two Majorana fermions with identical masses $\tilde{g}_{\mathrm{Dirac}} = \frac{1}{\sqrt{2}} \tilde{g}_1 + \frac{i}{\sqrt{2}} \tilde{g}_2$, the result is that Dirac rates are \emph{one half} of the rates in the Majorana case. This is seen because the gluino only couples through $\chi_O $. We have
\begin{align}
\Gamma (\tilde{g}_{\mathrm{Dirac}} \rightarrow g G)=&\frac{m_{D3}^5}{32\pi f^2}\,, \nn
\\
\Gamma (\tilde{g}_{\mathrm{Dirac}}\rightarrow q_j \ov{q}_j G)=& \frac{g_s^2 m_{D3}^5}{3072 \pi^3 f^2}\,.
\end{align} 

Finally, a particular case of interest  is Fake Split Supersymmetry \cite{Dudas:2013gga,Benakli:2013msa}, where $M_3 \gg m_{D}$, $M_{O} \sim \frac{m_{D}^2}{M_3}$. In this scenario, the lightest gluino eignenstate is mostly $\chi_O$ with small mixing between the two eigenstates given by $R_{11} \equiv \varepsilon \sim \frac{m_D}{M_3} \ll 1$. This then suppresses the couplings to the gravitino yields
\be
\Gamma (\tilde{g}_1 \rightarrow g G) = |\varepsilon|^2 \frac{m_{\tilde{g}_1}^5}{16 \pi f^2}\,,\qquad
\Gamma (\tilde{g}_1 \rightarrow q_j \ov{q}_j G) = |\varepsilon|^2 \frac{g_s^2  m_{\tilde{g}_1}^5}{1536 \pi^3 f^2}\,,
\ee
so that the corresponding lifetimes are longer than the standard ones of Split Supersymmetry; for large values of $M_3$ this could be a problem cosmologically \cite{Benakli:2013msa}.

We comment here that in the above analysis we have neglected the contribution from integrating out the octet scalar, which also couples to gluons and quarks (as discussed in section \ref{sgluondecays}). However, these couplings are suppressed by a loop factor in addition to the ratio $m_{D}/m_{\tilde{q}}^2$ (for the gluons) or $m_{D} m_t/m_{\tilde{q}}^2$ (for quarks) and are negligible relative to the tree-level ones above.

In summary, in Fake Split SUSY the decays of the gluino to a gravitino are parametrically suppressed relative to the Split SUSY case, something that is not a priori obvious, and which has consequences for how heavy the supersymmetry breaking scale can be.  

%%%%%%%%%%%%%%
\section{Conclusions}
%%%%%%%%%%%%%%

Motivated by the improvement that both low-scale SUSY breaking and Dirac gauginos bring to the naturalness problem as well as the indication that Dirac gaugino masses are actually \emph{necessary} for low-scale SUSY scenarios,  we studied the structure and the phenomenological implications of interactions between an ultralight gravitino and MSSM with Dirac gauginos.

By use of an effective field theory approach, we constructed the general expression of the couplings between the goldstino and the adjoint multiplets, extending earlier constructions for Majorana gauginos.

We then examined the phenomenological implications of these couplings. In agreement with earlier studies of generic goldstino couplings to matter, we showed that, in contrast to naive expectations, the new physical couplings do not include a coupling between two goldstinos and an electroweak gauge boson so that the Z-boson and photon coupling to goldstinos is not modified.

On the other hand, the decay pattern of the sgluon \emph{is} modified. In particular, the lighter of the two sgluon eigenstates can also decay to a gluon and two goldstinos. Since for a light enough sgluon the decay to two gluinos is kinematically forbidden, this novel process competes only with a loop-induced decay to two top quarks, leading to a significant modification of the decay pattern of the sgluon even for relatively high values of the SUSY breaking scale. Furthermore, since sgluons are dominantly pair produced, the latter process leads to the usual four-tops experimental signature while the former process to decays to two or three jets plus missing energy, which may be experimentally challenging -- i.e. in low-scale SUSY-breaking scenarios the sgluon may be even lighter than the presently weak bounds suggest. To illustrate the importance of these channels we computed the branching fractions of the sgluons for various scenarios and also computed the cross-section for such events at LHC13. In addition, in contrast to previous analyses of sgluons, we treated them as two sets of real scalars (since in Dirac gaugino models they are always split as such) rather than one complex scalar. 

The octet scalar may be produced individually through gluon fusion. If it subsequently decays to two gravitinos and a gluon then this may contribute to monojet signatures. However, we showed that in these scenarios the octet scalar and gaugino-mediated events leading to a monojet mutually interfere: if we completely integrate out the gluino, then, in fact, this decay channel will no longer be present. To illustrate the potential for discovery of such events at LHC13, we computed the cross-sections for two scenarios with low SUSY-breaking scale and light stops. In fact, there is a certain tension between increasing the sgluon production rate through gluon fusion and ensuring that gluons are not the dominant decay channel; however, we find that despite this, light octets can contribute a significant number of events. 

Finally, we examined how the gluino decay rate changes in the presence of light gravitinos, in the context of Split SUSY and Fake Split SUSY, where the squark masses are very high. The decay rates differ by a factor of two between a pure Majorana and a Dirac gluino, while the decay rate of the fake gluino in the Fake Split SUSY case leads to lifetimes parametrically larger than the standard Split SUSY case.

Clearly the phenomenological analyses described in this paper are just a first step, and we have merely illustrated the possibilities. We have already emphasised that it is now important to revisit existing constraints on low-energy supersymmetry-breaking scenarios with squarks and gauginos in the spectrum in terms of \emph{Dirac} gauginos. We expect the bounds on the supersymmetry-breaking scale to be weaker in this case. However, we are also proposing that an alternative simplified analysis would retain only the gluino and the sgluon, and we have provided all of the theoretical background to do this.

%%%%%%%%%%%%%%
\section*{Acknowledgements}
%%%%%%%%%%%%%%

We would both like to reserve special thanks to Emilian Dudas for collaboration and stimulating discussions during the earlier stages of this project. MDG would like to thank Marco Zaro for discussions on Madgraph, Pietro Slavich and Karim Benakli for interesting conversations. PT is supported in part by Vrije Universiteit Brussel through the Strategic Research Program `High-Energy Physics', in part by the Belgian Federal Science Policy Office through the Inter-university Attraction Pole P7/37 and in part by FWO-Vlaanderen through project G011410N. We thank the Galileo Galilei Institute for Theoretical Physics for the hospitality and the INFN for partial support in the initial stage of this work. This work was supported in part by the European ERC Advanced Grant 226371 MassTeV; MDG was also supported during part of this projet by the Marie-Curie contract no. PIEF-GA-2012-330060 BMM@3F.  

%%%%%%%%%%%%%%
\section{Appendix}
%%%%%%%%%%%%%%

\def\theequation{A-\arabic{equation}}
\def\thesubsection{A}
\setcounter{equation}{0}

%%%%%%%%%%%%%%
\subsection{Details of the effective action}
\label{Leff}
%%%%%%%%%%%%%%

Dirac gauginos are incorporated in the MSSM by adding to the field content one chiral superfield in the adjoint representation for every gauge group. The Lagrangian under study consists of the Nonlinear MSSM Lagrangian \cite{Antoniadis:2010hs} extended by operators that involve these new superfields.
\be\label{L}
\cL=\cL_{XMSSM}+\cL_{DG}\,.
\ee
It consists of
\bea\label{xmssm}
\cL_{XMSSM} &=&\!\!\sum_{\Phi} \int d^4\theta \left( 1-{m_\Phi^2\over f^2}X^\dag X \right) \Phi^\dagger\,e^{V_\Phi}\,\Phi+X^\dag X\nonumber
\\
&&\hspace{-2.5cm}+\bigg\{\int d^2\theta\,\Bigg[\left(\mu+{B\over f}X\right) H_1\cdot H_2+\left(y_u+{A_u\over f}X\right)H_2\cdot Q\,U^c+\left(y_d+{A_d\over f}X\right)Q\,D^c\cdot H_1\nonumber
\\
&&\hspace{-2.5cm}+\left(y_e+{A_e\over f}X\right)L\,E^c\cdot H_1+\sum_i\left( {1\over 4}+{M_i\over 2f}X \right) [W^{\alpha\,a}W_\alpha^a]_i+fX\Bigg]+h.c.\bigg\}\,,
\eea
where $ \Phi:H_1,H_2,Q,D^c,U^c,E^c,L$, $W^\alpha_{i=Y,w,c}$ is the field strength superfield for the hypercharge, the $SU(2)$ and the $SU(3)$ gauge fields respectively (e.g. $W_w=g_wW_w^a\sigma^a$) and $V_{H_1,H_2}=g_wV^i_w\sigma^i\mp g_YV_Y$. In the last line above, we have already taken the trace in the field strength superfields. Also,
\bea\label{ldg}
&&\hspace{-2.5cm}\cL_{DG}=\int d^4\theta\left(1-{m_{\Sigma_i}^2\over f^2}X^\dag X \right)2\,\Tr[\Sigma^{\dag}_ie^{V_i} \Sigma_i e^{-V_i}]
\\
&&\hspace{-2.5cm}+\bigg\{\int d^2\theta\,\Bigg[\left(\kappa_S+{A_S\over f}X\right) S H_1\cdot H_2+2\left(\kappa_T+{A_T\over f}X\right) H_1\cdot \overrightarrow{T}H_2\nonumber
\\
&&\hspace{-0.5cm}+ {1\over 2}\left({M_{\Sigma_i}+{B_i\over f}X}\right)[\Sigma^a\Sigma^a]_i-{\ov{D}^2D^\alpha(X^\dag X)\over 4\sqrt{2}f^2}m_{D_i}[ W_{\alpha}^a \Sigma^a]_i \Bigg]+h.c.\bigg\}\nonumber
\eea
where $V_i=g_YV_Y,\, g_wV_w^i\sigma^i,\, g_cV_c^a\lambda^a$ and  $\Sigma_i=S,\,\overrightarrow{T},\, \mathbf{O}$ are respectively the singlet superfield, the $SU(2)$-triplet superfield $\overrightarrow{T}=T^i\sigma^i/2$ and the $SU(3)$ octet $\mathbf{O}=O^a\lambda^a/2$, where $\sigma^{1,2,3}$ and $\lambda^{1,. . .,8}$ are the Pauli and Gell-Mann matrices. As before, in the last line above, we have already taken the trace in the field strength and the adjoint superfields. The symmetries of the theory allow for a linear superpotential term $m^2\,S$ as well as cubic interactions between the adjoint superfields, however in the present study we choose to neglect them.

The component expansion of (\ref{xmssm}) is given in \cite{Antoniadis:2010hs}. For completeness, we present the component expansion of certain parts that are of interest to this sudy. We have
\be
\int d^4\theta\, X^\dagger X=|\partial_\mu x|^2+F_X^\dag F_X +\Big(\frac{i}{2} \overline\psi_X\,\overline\sigma^\mu\partial_\mu\psi_X+h.c.\Big)\,,
\ee
\bea\label{XXHH}
&&\hspace{-0.7cm}\int d^4\theta \left(1-{m_j^2\over f^2}X^\dagger X\right) H_j^\dagger e^{V_j}H_j=\nonumber
\\
&=&
\vert \cD_\mu\, h_j\vert^2
+F_{j}^\dagger F_{j}+h_j^\dagger\,\frac{D_j}{2}\,h_j+
\Big(\frac{i}{2}\overline\psi_{j}\overline\sigma^\mu\,\cD_\mu\psi_{j}
-\frac{1}{\sqrt 2} \,h_j^\dagger \lambda_j\,\psi_{j}+h.c.\Big)\nonumber
\\
&-&{m_j^2\over f^2}\Bigg\{ |x|^2\,\Big[ \vert \cD_\mu\, h_j\vert^2
+F_{j}^\dagger F_{j}+h_j^\dagger\,\frac{D_j}{2}\,h_j+
\Big(\frac{i}{2}\overline\psi_{j}\overline\sigma^\mu\,\cD_\mu\psi_{j}
-\frac{1}{\sqrt 2} \,h_j^\dagger \lambda_j\,\psi_{j}+h.c.\Big)\Big]
\nonumber
\\
&+&\frac{1}{2}\, h_j^\dag\,(\cD_\mu+\overleftarrow\cD_\mu)\,h_j\partial^\mu |x|^2
+\overline\psi_X\overline\psi_{j}\,\psi_X\psi_{j} -\frac{1}{2}\,[x^\dag\,(\partial^\mu-\overleftarrow\partial^\mu)\,x]\,[ h_j^\dagger (\cD_\mu - \overleftarrow\cD_\mu)\,h_j]\nonumber
\\
&+&\Big[-\frac{i}{2}x^\dag \psi_X \sigma^\mu\overline\psi_{j} (\cD_\mu-\overleftarrow \cD_\mu) {h_j} -\frac{1}{\sqrt 2}\, x^\dag \psi_X h_j^\dag\lambda_j\,{h_j}-x^\dag \psi_X F_{j}^\dag\psi_{j}+x^\dag F_X\,F_{j}^\dagger {h_j}\nonumber
\\
&+&\frac{i}{2}(\overline\psi_X\overline\sigma^\mu\,\psi_X)(h_j^\dag\cD_\mu\,{h_j})
+\frac{i}{2}\,(x^\dag \partial_\mu x) (\overline\psi_{j}\,\overline\sigma^\mu\,\psi_{j})
+\frac{i}{2} \overline\psi_X\,\overline\sigma^\mu(\partial_\mu-\overleftarrow\partial_\mu)x (h_j^\dag \psi_{j})\nonumber
\\
&-& \overline\psi_X\,F_X\,\,\overline\psi_{j}\,{h_j}+h.c.\Big] +\Big[|\partial_\mu x|^2+F_X^\dag F_X +\Big(\frac{i}{2} \overline\psi_X\,\overline\sigma^\mu\partial_\mu\psi_X+h.c.\Big)\Big]
\vert\,{h_j}\vert^2\Bigg\},
\eea
where $j=1,\,2$ and $\cD^\mu h_j=(\partial^\mu+{i\over 2}V^\mu_j)h_j$ and $h_j^\dag\overleftarrow \cD^\mu=(\cD^\mu h_j)^\dag=h_j^\dag (\overleftarrow\partial^\mu-{i\over 2}V^\mu_j)$. The other Kahler operators are easily obtained upon appropriate substitution of the matter and gauge fields. For the adjoint superfields we need to take into account that for generic adjoint field $\phi$,
\be
V_\mu \phi\rightarrow [V_\mu,\phi];\quad\ \phi^\dag V_\mu\rightarrow [\phi^\dag,V_\mu],\ \textrm{same for $\lambda$ and D}\,,
\ee
for example,
\begin{align}\label{XXOO}
\int d^4\theta &\left(1-{m_O^2\over f^2}X^\dagger X\right) 2 \tr (\mathbf{O}^\dagger e^{V}\mathbf{O}e^{-V}) =\nonumber
\\
=&
\vert \cD_\mu\, O^a\vert^2
+F_{O}^\dagger F_{O}+ \tr (\ov{O} [ D_3,O])+
\Big(\frac{i}{2}\overline\chi_O^a\overline\sigma^\mu\,\cD_\mu\chi_O^a
-\frac{1}{\sqrt 2} \,2 \tr(\ov{O} [\tilde{g},\chi_O])+h.c.\Big)\nonumber
\\
-&{m_O^2\over f^2}\Bigg\{ |x|^2\,\Big[ \vert \cD_\mu\,O^a\vert^2
+ F_{O}^\dagger F_{O} +\tr (\ov{O} [ D_3,O])+
\Big(\frac{i}{2}\overline\chi_O^a\overline\sigma^\mu\,\cD_\mu\chi_O^a
-\frac{1}{\sqrt 2} \,\ov{O}^a \tilde{g}^a\,\chi_O^a+h.c.\Big)\Big]
\nonumber
\\
+&\frac{1}{2}\, \ov{O}^a\,(\cD_\mu+\overleftarrow\cD_\mu)\,O^a\partial^\mu |x|^2
+\overline\psi_X\overline\chi_O^a\,\psi_X\chi_O^a -\frac{1}{2}\,[x^\dag\,(\partial^\mu-\overleftarrow\partial^\mu)\,x]\,[ \ov{O}^a (\cD_\mu - \overleftarrow\cD_\mu)\,O^a]\nonumber
\\
+&\Big[-\frac{i}{2}x^\dag \psi_X \sigma^\mu\overline\chi_O^a (\cD_\mu-\overleftarrow \cD_\mu) {O^a} -\frac{1}{\sqrt 2}\, x^\dag \psi_X \ov{O}^a\tilde{g}^a\,{O^a}-x^\dag \psi_X F_{O}^\dag\chi_O^a+x^\dag F_X\,F_{O}^\dagger {O^a}\nonumber
\\
+&\frac{i}{2}(\overline\psi_X\overline\sigma^\mu\,\psi_X)(\ov{O}^a\cD_\mu\,{O^a})
+\frac{i}{2}\,(x^\dag \partial_\mu x) (\overline\chi_O^a\,\overline\sigma^\mu\,\chi_O^a)
+\frac{i}{2} \overline\psi_X\,\overline\sigma^\mu(\partial_\mu-\overleftarrow\partial_\mu)x (\ov{O}^a \chi_O^a)\nonumber
\\
-& \overline\psi_X\,F_X\,\,\overline\chi_O^a\,{O^a}+h.c.\Big] +\Big[|\partial_\mu x|^2+F_X^\dag F_X +\Big(\frac{i}{2} \overline\psi_X\,\overline\sigma^\mu\partial_\mu\psi_X+h.c.\Big)\Big]
\vert\,{O^a}\vert^2\Bigg\},
\end{align}

Also,
\bea\label{XWW}
&&\hspace{-1cm}\int d^2\theta \Bigg[\left( {1\over 4}+{M_i\over 2f}X \right) [W^{\alpha\,a}W_\alpha^a]_i-{\ov{D}^2D^\alpha(X^\dag X)\over 4\sqrt{2}f^2}m_{D_i}[ W_{\alpha}^a \Sigma^a]_i\Bigg]+h.c.=\nonumber
\\
&&\hspace{-1cm}=\Bigg[-{1\over 4}F_{\mu\nu}^aF^{\mu\nu\,a}+i\lambda^a \sigma^\mu(\Delta_\mu\ov{\lambda})^a+{D^aD^a\over 2}\Bigg]_i\nonumber
\\
&&\hspace{-1cm}+\,{M_i\over 2f}\Bigg[ x\left( -{1\over 2}F^{\mu\nu\,a}F_{\mu\nu}^a+2i\lambda^a\sigma^\mu(\Delta_\mu\ov{\lambda})^a+D^aD^a-{i\over 4}\epsilon^{\mu\nu\rho\sigma}F^a_{\mu\nu}F_{\rho\sigma}^a \right)\nonumber
\\
&&\hspace{0.5cm}+\sqrt{2}\lambda^a\sigma^{\mu\nu}\psi_XF_{\mu\nu}^a-\sqrt{2}\psi_X\lambda^a D^a+F_X\lambda^a\lambda^a +h.c.\Bigg]_i\nn
\\
&&\hspace{-1cm}-\,{m_{D_i}\over \sqrt{2}f^2}\Bigg[ -\sqrt{2}F_X^\dag \psi_X\lambda^a F^a+D^aF_X^\dag \psi_X\chi^a-\sqrt{2}iF_X^\dag \psi_X\sigma^\mu(\Delta_\mu\ov{\lambda})^as^a-F_X^\dag \psi_X\sigma^{\mu\nu}F_{\mu\nu}^a\chi^a\nonumber
\\[2pt]
&&\hspace{0.6cm}+\sqrt{2}|F_X|^2\lambda^a\chi^a-2|F_X|^2D^as^a-\sqrt{2}i\chi^a\sigma^\mu\partial_\mu(\ov{\psi}_X\psi_X)\lambda^a-i\partial_\mu(\psi_X\sigma^\mu\ov{\psi}_X)D^as^a\nonumber
\\[2pt]
&&\hspace{0.6cm}+i\partial_\rho(\psi_X\sigma^{\mu\nu}\sigma^\rho \ov{\psi}_X)F_{\mu\nu}^as^a+\sqrt{2}i\partial_\mu(\ov{\psi}_X\ov{\sigma}^\mu F_X)\lambda^a s^a-\sqrt{2}i\partial_\mu \psi_X\sigma^\mu\ov{\psi}_X\lambda^a\chi^a\nonumber
\\[2pt]
&&\hspace{0.6cm}+2i\partial_\mu \psi_X \sigma^\mu\ov{\psi}_XD^as^a+h.c.\Bigg]_i
\eea
where $s_i$ is the scalar field in $\Sigma_i$, $(\Delta_\mu\ov{\lambda})^a=\partial_\mu\ov{\lambda}^a-gt^{abc}V_\mu^b\ov{\lambda}^c$ and $F_{\mu\nu}^a=\partial_\mu V_\nu^a-\partial_\nu V_\mu^a-gt^{abc}V_\mu^bV_\nu^c$ (for the hypercharge we have the abelian expressions - for the $SU(2)$ gauge group, $t^{abc}=\epsilon^{abc}$). Notice that the ``Dirac gaugino" operator does not contain 2-goldstino couplings coming from sgoldstino exchange. Finally,
\bea\label{Leff9}
&&\hspace{-1cm}\int d^2\theta\Bigg[\!\!\left(\!\kappa_S\!+\!{A_S\over f}X\!\right)\! S H_1\!\cdot\! H_2+2\!\left(\!\kappa_T\!+\!{A_T\over f}X\!\right)\!\! H_1\!\cdot\! \overrightarrow{T}H_2+ {1\over 2}\!\left(\!M_{\Sigma_i}\!+\!{B_i\over f}X\!\right)\![\Sigma^a\Sigma^a]_i\Bigg]\!+\!h.c.\nonumber
\\
&&\hspace{-1cm}=(k_S+{A_S\over f}x)\Big[ s(h_1\!\cdot\!F_2+F_1\!\cdot\!h_2-\psi_1\!\cdot\!\psi_2)-\chi_S(\psi_1\!\cdot\!h_2+h_1\!\cdot\!\psi_2)+F_Sh_1\!\cdot\!h_2 \Big]\nonumber
\\
&&\hspace{-1cm}+\,(k_T+{A_T\over f}x)\Big[ h_1\!\cdot\!(F_T^i\sigma^i h_2+s_T^i\sigma^i F_2-\chi_T^i\sigma^i\psi_2)-\psi_1\!\cdot\!(\chi_T^i\sigma^i h_2+s_T^i\sigma^i \psi_2)+F_1\!\cdot\! s_T^i\sigma^i h_2 \Big]\nonumber
\\
&&\hspace{-1cm}+\,(M_{S}+{B_S\over f}x)\Big(F_Ss-{1\over 2}\chi_S\chi_S\Big)+(M_{T}+{B_T\over f}x)\Big(F_T^is_T^i-{1\over 2}\chi_T^i\chi_T^i\Big)\nonumber
\\
&&\hspace{-1cm}+(M_{O}+{B_O\over f}x)\Big(F_O^aO^a-{1\over 2}\chi_O^a\chi_O^a\Big)+{F_X\over f}\left[A_Ssh_1\!\cdot\!h_2+A_Th_1\!\cdot\! s_T^i\sigma^ih_2+{B_i\over 2}(s^as^a)_i\right]\nonumber
\\
&&\hspace{-1cm}\,-{\psi_X\over f}\Big[ A_S(\chi_Sh_1\!\cdot\!h_2+s\psi_1\!\cdot\!h_2+sh_1\!\cdot\!\psi_2)+A_T(\psi_1\!\cdot\! s_T^a\sigma^ah_2+h_1\!\cdot\!\chi_T^a\sigma^ah_2+h_1\!\cdot\! s^a_T\sigma^a\psi_2)\nonumber
\\
&&\hspace{0.2cm}+B_i(\chi^as^a)_i \Big]+h.c.
\eea

The terms that couple a sgluon to a gluino and a goldstino (including terms after integrating out the auxiliary fields) are given by
\bea
\cL_{O\tilde{g}G}&=& -{B_O\over f}O^a\psi_X\chi_O^a-{m_{O}^2\over f}O^a\ov{\chi}_O^a\ov{\psi}_X-{m_{D3}^2\over f}(O^a+O^{a*})\psi_X\chi_O^a\nonumber
\\
&+&{m_{D3}M_{O}\over f}O^{a*}\psi_X\lambda_c^a+{m_{D3}M_3\over f}(O^a+O^{a*})\psi_X\lambda_c^a-{m_{D3}\over f}i\psi_X\sigma^\mu(\Delta_\mu \ov{\lambda}_c)^aO^a+h.c.\nonumber\,,
\eea
while the coupling of a sgluon to a gluon and two goldstinos is
\be
\cL_{OgGG}= -{m_{D3}\over \sqrt{2}f^2}i\partial_\rho(\psi_X\sigma^{\mu\nu}\sigma^\rho\ov{\psi}_X)G^a_{\mu\nu}O^a+h.c.
\ee

\def\theequation{B-\arabic{equation}}
\def\thesubsection{B}
\setcounter{equation}{0}

%%%%%%%%%%%%%%%
\subsection{Goldstino - neutralino mixing}
%%%%%%%%%%%%%%%

The minimum of the scalar potential of the CP-even neutral fields is given by
\be
V=  F_S^2 + F_T^2 + F_1^2 + F_2^2 + {D_1^2 \over 2} + {D_2^2\over 2} + V_{\textrm{soft}}
\ee
where
\bea
F_S &=& -k_S {v_1v_2\over 2} - M_S {v_S\over \sqrt{2}} \nn
\\
F_T &=& k_T {v_1 v_2\over 2}- M_T {v_T\over\sqrt{2}}\nn
\\
F_1 &=& -(\mu + k_S {v_S\over \sqrt{2}} - k_T {v_T\over \sqrt{2}}) {v_2\over \sqrt{2}}\nn
\\
F_2 &=& (\mu + k_S {v_S\over \sqrt{2}} - k_T {v_T\over \sqrt{2}}) {v_1\over \sqrt{2}}\nn
\\
D_1 &=& {g_1\over 4} (v_1^2 - v_2^2) - 2 m_{D1} v_S\nn
\\
D_2 &=& -{g_2\over 4} (v_1^2 - v_2^2) - 2 m_{D2} v_T
\eea
and
\be
V_{\textrm{soft}} = m_1^2 {v_1^2\over 2} + m_2^2 {v_2^2\over 2} + B v_1 v_2 + m_S^2 {v_S^2\over 2} + B_S {v_S^2\over 2 }+ m_T^2 {v_T^2\over 2} + B_T {v_T^2\over 2} + A_S v_S {v_1 v_2\over \sqrt{2}} - A_T v_T {v_1 v_2\over \sqrt{2}}
\ee
The minimisation conditions $\partial V/ \partial v_I=0\,,\ I=1,2,S,T$ are used to express four parameters in terms of the others and will be used in the following. Also, the Dirac mass operator and electroweak symmetry breaking modify the kinetic terms, introducing mixing between the goldstino and the gauginos. In particular
\bea
\cL_k&=&i\lambda_Y\sigma^\mu\partial_\mu \ov{\lambda}_Y+i\lambda_w^3\sigma^\mu\partial_\mu \ov{\lambda}_w^3\nn
\\
&&+i\left( 1-{m_1^2v_1^2\over 2f^2}-{m_2^2v_2^2\over 2f^2}+{2 v_Sm_{D1} D_1\over f^2}+{2 v_Tm_{D2} D_2\over f^2} \right)\psi_X\sigma^\mu\partial_\mu \ov{\psi}_X\nonumber
\\
&&-{\sqrt{2}i\over f}\left[ m_{D1}v_S(\psi_X\sigma^\mu\partial_\mu\ov{\lambda}_Y+\lambda_Y\sigma^\mu\partial_\mu\ov{\psi}_X)+m_{D2}v_T^3(\psi_X\sigma^\mu\partial_\mu\ov{\lambda}_w^3+\lambda_w^3\sigma^\mu\partial_\mu\ov{\psi}_X) \right]\,.\nonumber
\eea
The kinetic terms are diagonalised by performing the following field redefinitions
\be\label{diag1}
\lambda_Y\rightarrow\lambda_Y+{\sqrt{2}m_{D1}\over f}v_S\psi_X\,,\quad \lambda_w^3\rightarrow\lambda_w^3+{\sqrt{2}m_{D2}\over f}v_T\psi_X\,.
\ee
followed by a rescaling of the goldstino field
\be\label{diag2}
\psi_X\to{\psi_X\over \sqrt{\xi}}
\ee
with
\be
\xi= 1-{1\over 2f^2}\left( 12v_S^2m_{D1}^2+12v_T^2m_{D2}^2+m_1^2v_1^2+m_2^2v_2^2+(g_1m_{D1}v_S-g_2m_{D2}v_T)(v_2^2-v_1^2) \right)\,.
\ee
The $7\times 7$ neutralino mass matrix is given by 
\be
\cL_{neutr}=-{1\over 2}\cM_{lm}\Psi_l\Psi_m+h.c.
\ee
where $\Psi=(\psi_X\,\,\psi_1^0\,\,\psi_2^0\,\,\lambda_Y\,\,\lambda_w^3\,\,\chi_Y\,\,\chi_w^3)^T$ and $M_{lm}=M_{ml}$. The mass mixing terms, after taking into account the diagonalisation of the kinetic terms, are given by
\bea
\cM_{\psi_X\psi_X}&=&{2\over f^2}\bigg[m_{D1}(M_1v_SD_1+M_1m_{D1}v_S^2+\sqrt{2}m_{D1}v_SF_S)+{M_1\over 4}<\!\!D_1\!\!>^2\nonumber
\\
&+&m_{D2}(M_2v_T^3D_2^3+M_2m_{D2}(v_T^{3\,})^2+\sqrt{2}m_{D2}v_T^3F_T^3)+{M_2\over 4}<\!\!D_2^3\!\!>^2\nonumber
\\
&-&{1\over 4\sqrt{2}}\bigg(m_1^2<\!\!F_1\!\!>v_1+m_2^2<\!\!F_2\!\!>v_2+m_S^2<\!\!F_S\!\!>v_S+m_T^2<\!\!F_T^3\!\!>v_T^3\bigg) \nonumber
\\
&+&\!\!\!{A_S\over 4}\big[v_S(v_1<\!\!F_2^0\!\!>+<\!\!F_1^0\!\!>v_2)+<\!\!F_S\!\!>v_1v_2\big]\nn
\\
&-&{A_T\over 4}\big[v_T^3(v_1<\!\!F_2^0\!\!>+<\!\!F_1^0\!\!>v_2)+<\!\!F_T^3\!\!>v_1v_2\big]\nonumber
\\
&+&{B_S\over 2\sqrt{2}}<\!\!F_S\!\!>v_S+{B_T\over 2\sqrt{2}}<\!\!F_T^3\!\!>v_T^3+{B\over 2\sqrt{2}}(<\!\! F_1^0\!\!>v_2+v_1\!\!<\!\!F_2^0\!\!>)\bigg]\nn
\eea
\bea
\cM_{\psi_X\psi_1^0}&\!\!\!\!\!=&\!\!\!\!\!-{1\over \sqrt{2}f}\left(-Bv_2+ m_{D1}g_Yv_1v_S-m_{D2}g_wv_1v_T^3-m_1^2v_1 \right)+{v_2\over 2f}(A_Sv_S-A_Tv_T^3)\nonumber
\\
\cM_{\psi_X\psi_2^0}&\!\!\!\!\!=&\!\!\!\!\!-{1\over \sqrt{2}f}\left(-Bv_1 -m_{D1}g_Yv_2v_S+m_{D2}g_wv_2v_T^3-m_2^2v_2 \right)+{v_1\over 2f}(A_Sv_S-A_Tv_T^3)\nonumber
\\
\cM_{\psi_X\lambda_Y}&\!\!\!\!\!=&\!\!\!\!\!{1\over \sqrt{2}f}\left( M_1<\!\!D_1\!\!>+2M_1m_{D1}v_S+\sqrt{2}m_{D1}<\!\!F_S\!\!> \right)\nonumber
\\
\cM_{\psi_X\lambda_w^3}&\!\!\!\!\!=&\!\!\!\!\!{1\over \sqrt{2}f}\left( M_2<\!\!D_2^3\!\!>+2M_2m_{D2}v_T^3+\sqrt{2}m_{D2}<\!\!F_T^3\!\!> \right)\nonumber
\\
\cM_{\psi_X\chi_Y}&\!\!\!\!\!=&\!\!\!\!\!{1\over \sqrt{2}f}\left( m_S^2v_S-m_{D1}<\!\!D_1\!\!>+2m_{D1}^2v_S+{A_S\over \sqrt{2}}v_1v_2+B_Sv_S\right)\nonumber
\\
\cM_{\psi_X\chi_w^3}&\!\!\!\!\!=&\!\!\!\!\!{1\over \sqrt{2}f}\left( m_T^2v_T^3-m_{D2}<\!\!D_2^3\!\!>+2m_{D2}^2v_T^3-{A_T\over \sqrt{2}}v_1v_2+B_Tv_T^3 \right)\nonumber
\eea
\vspace{-1cm}
\bea
&&\cM_{\psi_1^0\psi_2^0}=\mu+{\left( \kappa_Sv_S-\kappa_Tv_T^3 \right)\over \sqrt{2}},
\cM_{\psi_1^0\lambda_Y}=-{v_1g_Y\over 2},
\cM_{\psi_1^0\lambda_w^3}={v_1g_w\over 2},\nn
\\
&&\cM_{\psi_1^0\chi_Y}={\kappa_Sv_2\over \sqrt{2}},\,,\cM_{\psi_1^0\chi_w^3}=-{\kappa_Tv_2\over \sqrt{2}}\nonumber
\eea
\be
\cM_{\psi_2^0\lambda_Y}={1\over 2}v_2g_Y\,,\quad
\cM_{\psi_2^0\lambda_w^3}=-{1\over 2}v_2g_w\,,\quad
\cM_{\psi_2^0\chi_Y}={v_1\kappa_S\over \sqrt{2}}\,,\quad
\cM_{\psi_2^0\chi_w^3}=-{v_1\kappa_T\over \sqrt{2}}\nonumber
\ee
\be
\cM_{\lambda_Y\lambda_Y}=(1+{1\over 2f^2}(Bv_1v_2+m_1^2v_1^2+m_2^2v_2^2+m_S^2v_S^2+m_T^2(v_T^{3})^2))M_1\nn
\ee
\be
\cM_{\lambda_w^3\lambda_w^3}=(1+{1\over 2f^2}(Bv_1v_2+m_1^2v_1^2+m_2^2v_2^2+m_S^2v_S^2+m_T^2(v_T^{3})^2))M_2\nn
\ee
\be
\cM_{\lambda_Y\chi_Y}=(1+{1\over f^2}(Bv_1v_2+m_1^2v_1^2+m_2^2v_2^2+m_S^2v_S^2+m_T^2(v_T^{3})^2))m_{D1}\nn
\ee
\be
\cM_{\lambda_w^3\chi_w^3}=(1+{1\over f^2}(Bv_1v_2+m_1^2v_1^2+m_2^2v_2^2+m_S^2v_S^2+m_T^2(v_T^{3})^2))m_{D2}\nonumber
\ee
\be
\cM_{\chi_Y\chi_Y}=M_S \,,\quad
\cM_{\chi_w^3\chi_w^3}=M_T\nonumber
\ee
\be
\cM_{\chi_Y\chi_w^3}=\,
\cM_{\chi_Y\chi_w^3}=\,
\cM_{\lambda_w^3\chi_Y}=\,
\cM_{\lambda_Y\lambda_w^3}=\,
\cM_{\lambda_Y\chi_w^3}=\,
\cM_{\psi_2^0\psi_2^0}=\,
\cM_{\psi_1^0\psi_1^0}=0\nn
\ee
The goldstino direction in each fermion is identified by solving the linear system
\be
\cM_{lm}\Psi_m=0\,.
\ee
After simplifying the result by use of the minimisation conditions, we finally obtain
at lowest order in $f^{-1}$ the simple relations:
\bea\label{goldmix}
&&\psi_1={\tilde{\mu} v_2\over \sqrt{2}f}G+...\,,\quad \psi_2={\tilde{\mu} v_1\over \sqrt{2}f}G+...\nn
\\
&&\lambda_Y=-{g_1(v_1^2-v_2^2)\over 4\sqrt{2}f}G+...\,,\quad \lambda_w^3={g_2(v_1^2-v_2^2)\over 4\sqrt{2}f}G+...\nn
\\
&&\chi_S={\sqrt{2}M_Sv_S+k_Sv_1v_2\over 2f}G+...\,,\quad \chi_T^3={\sqrt{2}M_Tv_T-k_Tv_1v_2\over 2f}G+...
\eea
where $\tilde{\mu}=\mu+(k_Sv_S-k_Tv_T)/\sqrt{2}$ is the modified $\mu$ term and $G$ is the true goldstino (the massless eigenvector of the neutralino mass matrix).

%%%%%%%%%%%%%%
%%%%%%%%%%%%%%

\end{document}